\documentclass{aa}  

\usepackage{graphicx}
\usepackage{txfonts}
\usepackage{subcaption}
\usepackage{ulem}
\usepackage{lineno}

\usepackage{xcolor}

\newcommand{\msun}{${\rm M}_\odot$}
\newcommand{\lsun}{${\rm L}_\odot$}
\newcommand{\rsun}{${\rm R}_\odot$}
\newcommand{\msunyr}{${\rm M}_\odot {\rm yr}^{-1}$}
\def\deg{$^\circ$}

\def\kms{km\,s$^{-1}$}

\def\teff{$T_{\rm eff}$}
\def\logg{$\log g$}
\def\vsini{$v\sin i$}
\def\prot{${\rm P}_{rot}$}
\def\vrad{${\rm V}_{r}$}
\def\vmic{${\rm V}_{mic}$}
\def\vmac{${\rm V}_{mac}$}

\def\ha{${\rm H}\alpha$}
\def\hb{${\rm H}\beta$}
\def\hg{${\rm H}\gamma$}
\def\He~I{${\rm He~I}$}

\def\Ca~II{Ca~II}
\def\macc{$\dot{M}_{\rm acc}$}

\def\rstar{${\rm R}_\star$}
\def\mstar{${\rm M}_\star$}
\def\lstar{${\rm L}_\star$}
\def\phirot{$\Phi_{rot}$}
\def\rcor{$r_{cor}$}

\def\rmag{$r_{mag}$}

\begin{document} 

   \title{Beyond the dips of V807 Tau, a spectropolarimetric study of a dipper's magnetosphere\thanks{Based on observations obtained at the Canada-France-Hawaii Telescope (CFHT), which is operated by the National Research Council of Canada, the Institut National des Sciences de l'Univers of the Centre National de la Recherche Scientifique of France, and the University of Hawaii.}}

   \author{K.~Pouilly \inst{1}
          \and
          J. Bouvier \inst{1}
          \and
          E.~Alecian\inst{1}
          \and
          S.H.P.~Alencar\inst{2}
          \and
          A.-M.~Cody\inst{3}
          \and
          J.-F.~Donati\inst{4}
          \and
          K.~Grankin\inst{5}
           \and
          L.~Rebull\inst{6}
          \and
          C.P.~Folsom\inst{7,8}
          }
    
   \institute{Univ. Grenoble Alpes, CNRS, IPAG, 38000 Grenoble, France\\
              \email{kim.pouilly@univ-grenoble-alpes.fr}
              \and
              Departamento  de  Fisica - ICEx - UFMG,  Av.  Ant\^onio  Carlos  6627,  30270-901  Belo  Horizonte,  MG,  Brazil
              \and
              SETI Institute, 189 N Bernardo Ave \#200, Mountain View, CA 94043 USA
              \and
              Univ. de Toulouse, CNRS, IRAP, 14 avenue Belin, 31400 Toulouse, France
              \and
              Crimean Astrophysical Observatory, Nauchny, Crimea 298409
              \and
              Infrared Science Archive (IRSA), IPAC, California Institute of Technology, 1200 E. California Blvd, MS 100-22, Pasadena, CA 91125 USA
              \and
              Department of Physics \& Space Science, Royal Military College of Canada, PO Box 17000 Station Forces, Kingston, ON, Canada K7K 0C6
              \and
              Tartu Observatory, University of Tartu, Observatooriumi 1, Tõravere, 61602 Tartumaa, Estonia
                }

   \date{Received March 22, 2021; accepted September 13, 2021}

 
  \abstract
   {The so-called dippers are pre-main-sequence objects that accrete material from their circumstellar disks through the stellar magnetosphere. Their unique type of variability allows us to probe the magnetic star-disk interaction processes in young stellar objects.}
   {We aim to characterize the magnetospheric accretion process in the young stellar object V807 Tau, one of the most stable dippers revealed by \textit{K2} in the Taurus star forming region. }
   {We performed photometric and spectropolarimetric follow-up observations of this system with CFHT/ESPaDOnS in order to investigate the variability of the system over several rotational periods. }
   {We derive a 4.38 day period from the \textit{K2} dipper light curve. This period is also seen in the radial velocity variations, which we ascribe to spot modulation. The slightly redshifted narrow component of the He~I 5876 \AA\ line as well as the high velocity red wing of the H$\beta$ and H$\gamma$ emission line profiles also vary in intensity with the same periodicity. The former traces the accretion shock at the stellar surface, and the latter is a signature of an accretion funnel flow crossing the line of sight. We derive a surface brightness map and the topology of the surface magnetic field from the modeling of Stokes I and V profiles, respectively, for photospheric lines and for the He~I emission line. The latter reveals a bright spot at the stellar surface, located at a latitude of  60\deg, and a maximum field strength of $\sim$2 kG at this location. The topology of the magnetic field at the stellar surface is dominated by a dipolar component inclined by about 40$\degr$ onto the spin axis. Variable blueshifted absorption components seen in the Balmer line profiles suggest episodic outflows. Despite of its clear and stable dipper behavior, we derive a relatively low inclination of 40$\degr$ to 50$\degr$ for this system, which calls question the origin of the dips. The low inclination we infer is also consistent with the absence of deep inverse P Cygni components in the line profiles. }
   {We conclude that magnetospheric accretion is ongoing in V807 Tau, taking place through non-axisymmetric accretion funnel flows controlled by a strong, tilted, and mainly dipolar magnetic topology. Whether an inner disk warp resulting from this process can account for the dipper character of this source remains to be seen, given the low inclination of the system.}


   \maketitle

%

\section{Introduction}
    The study of classical T Tauri stars (CTTSs) provides important constraints on the evolution of young stellar objects.
    These low-mass stars actively accrete material from their circumstellar disks and are characterized by strong photometric and spectral variability, part of which arises from the star-disk interaction process. 
    Their strong surface magnetic field is able to truncate the inner disk, leading to  accretion funnel flows controlled by the magnetic field lines \citep{Bouvier07b, Hartmann16}.
    The magnetospheric star-disk interaction takes place on a scale of $\leq$0.1 au, which cannot be resolved by direct imaging.


    A common approach for investigating the star-disk interaction process is to monitor the system over several rotational periods,  using photometry, spectroscopy, spectropolarimetry, and, more recently, interferometry \citep[e.g.,][] {Bouvier07a,Donati10, Donati19, Alencar12, Alencar18, Pouilly20, Bouvier20b, Bouvier20a}. 
    Such studies have revealed strong magnetic fields, on the order of 1 kG at the stellar surface, often dominated by a dipolar component, as well as cool magnetic and hot accretion spots close to the magnetic poles \citep{Donati11, Donati12, Donati20}. 


    Some CTTSs show periodic dimming events in their light curves, referred to as "dips." The so-called dippers are thought to be systems seen at high inclination, with an inner disk warp periodically occulting the central star \citep[e.g.,][]{Bouvier99, Bouvier03, Alencar10, Mcginnis15}. The disk warp, in turn, results from the inner disk interacting with an inclined stellar magnetosphere \citep{Terquem00}. The warp is thus located at a few stellar radii above the stellar surface and corresponds to the base of the accretion funnel flow that connects the inner disk edge to the star. The dips seen in the light curves occur at the stellar rotation period, which suggests that the inner disk warp is  located close to the corotation radius. 

    This study focuses on \object{V807 Tau} (RA = 04h 33m Dec = +24\deg 09'), a member of the nearby Taurus star forming region. This multiple system, with a semimajor axis of $\sim$34 au \citep{Simon96}, consists of a K7 CTTS primary \citep{Nguyen12,Herczeg14} and a non-accreting companion, itself a binary consisting of two M-type stars with an orbital period of $\sim$ 12 yr and a semimajor axis of $\sim$ 4.5 au \citep{Schaefer12}.
    In the following, we use the name V807 Tau to refer to the primary only, a low-mass T Tauri star (TTS) (0.7 \msun) located at 113pc \citep{Gaia18}, for which 
    \cite{Schaefer12} report a radial velocity <\vrad> = 15.79 $\pm$ 0.65 \kms and a projected rotational velocity \vsini\,=\,15\,\kms. 
    V807 Tau was identified as a dipper by \cite{Rodriguez17} from ground-based photometric monitoring, and \cite{Rebull20} derived a photometric period of 4.3784 days from its 80-day-long \textit{K2} light curve. 
    
    The remarkable periodic dipper behavior of V807 Tau revealed by the \textit{K2} campaign on Taurus motivated us to follow up on this system with high-resolution spectropolarimetry at CFHT.
    The dips of the K2 light curve are very stable and well defined, allowing us to constrain the period of the occulting dust and compare it with the stellar rotation period derived from spectroscopy.
    Furthermore, the periodicity of the dips is relatively short, offering the opportunity to cover a larger number of rotational cycles within an observing run.
    The main goal was to derive the magnetic topology at the stellar surface and probe the star-disk  interaction region in order to investigate the origin of the periodic photometric dips.

    In Section \ref{sec:obs}, we present the different data sets we used for this study.
    In Section \ref{sec:results}, we present the results obtained from the photometric, spectroscopic, and spectropolarimetric monitoring of the system. We derive the stellar parameters of the system, analyze the variability of various emission lines present in its spectrum, and derive surface brightness and magnetic maps. 
    We discuss the results in the framework of the magnetospheric accretion process in Section \ref{sec:discussion} and present the main conclusions in Section \ref{sec:conclusion}.


\section{Observations}
\label{sec:obs}
    We describe in this section the acquisition and reduction of the photometric and spectropolarimetric data-sets used in this work.

    \subsection{Photometry}
        The \textit{K2} light curve of V807 Tau was obtained during Campaign 13 on Taurus and extends over  80 days, from March 8 to May 27, 2017. 
        The measurements were performed through a broad band filter (420-900 nm) at a cadence of 30 minutes.
        In this work we used the pre-search data conditioning (PDC) (i.e., corrected for the instrumental variations) version of the light curve \citep{Cody18}. This photometry was obtained about 6 months before our spectroscopic follow-up campaign. 

    Multicolor photometry was secured at the Crimean Astrophysical Observatory (CrAO) contemporaneously with the CFHT/ESPaDOnS run. 
    Photometric measurements were collected in the VR$_j$I$_j$ bands from October 12, 2017, to March 16, 2018, on the AZT-11 1.25m telescope equipped with the charge-coupled device (CCD) camera ProLine PL23042. 
    The CCD images were bias subtracted and flat-field corrected following a standard procedure. We performed differential photometry between V807 Tau and a non-variable, nearby comparison star, Gaia DR2 147589271158748928, whose brightness and colors in the Johnson system are V=14.12, (V-R)$_j$=1.69, and (V-I)$_j$=3.18.
    A nearby control star of similar brightness, Gaia DR2 147588687043206784, was used to verify
    that the comparison star was not variable. 
    It also provides an estimate of the photometric rms error in each filter, which amounts to 0.013, 0.015, and 0.014 in the VR$_j$I$_j$ bands, respectively.
    Table~\ref{tab:craophot} lists the V, (V-R)$_j$, and (V-I)$_j$ measurements. 
    Even though the photometric data set spans over 150 days, from JD 8039 to JD 8194, the sampling is sparse, with a lack of measurements obtained simultaneously with the spectroscopic observations.

        \begin{table}
            \centering
            \begin{tabular}{l l l l}
                \hline
                HJD & V & (V-R)$_J$ & (V-I)$_J$ \\
                (2,450,000+) \\
                \hline
                8039.62 & 11.567 & 1.230 & 2.402\\
                8043.57 & 11.480 & 1.217 & 2.354\\
                8044.54 & 11.512 & 1.212 & 2.371\\
                8045.52 & 11.489 & 1.207 & 2.358\\
                8069.59 & 11.569 & 1.247 & 2.382\\
                8082.57 & 11.515 & 1.183 & 2.349\\
                8083.54 & 11.524 & 1.203 & 2.372\\
                8101.29 & 11.513 & 1.222 & 2.381\\
                8114.37 & 11.606 & 1.232 & 2.42\\
                8115.37 & 11.541 & 1.233 & 2.401\\
                8125.40 & 11.505 & 1.227 & 2.373\\
                8126.24 & 11.468 & 1.208 & 2.344\\
                8128.30 & 11.514 & 1.236 & 2.379\\
                8144.23 & 11.691 & 1.241 & 2.459\\
                8157.22 & 11.772 & 1.295 & 2.519\\
                8165.28 & 11.486 & 1.221 & 2.385\\
                8171.23 & 11.665 & 1.277 & 2.472\\
                8183.24 & 11.548 & 1.224 & 2.413\\
                8194.22 & 11.530 & 1.246 & 2.409\\
            \hline\end{tabular}
            \caption{CrAO photometry obtained for the V807 Tau system in the V-band and (V-R)$_J$, (V-I)$_J$ colors.}
            \label{tab:craophot}
        \end{table}
        
    \subsection{Spectropolarimetry}
        We obtained high-resolution spectropolarimetry using the Echelle SpectroPolarimetric Device for the Observation of Stars (ESPaDOnS) \citep{Donati03}.
        ESPaDOnS is mounted at the Canada-France-Hawaii Telescope (CFHT) and allows spectropolarimetric observation in the 370 - 1000 nm range at a resolution of 68,000.
        The observations are composed of four different spectra in different polarization states.
        Each observation is reduced using the Libre-ESpRIT package \citep{Donati97} and provides an unpolarized (Stokes I) and a circularly polarized (Stokes V) spectrum.
        The V807 Tau campaign took place between October 28 and November 9, 2017, with a sampling rate of approximately 1 spectrum per night.
        The 11 observations reach a signal-to-noise ratio (S/N) between 140 and 190 at 731 nm. Spectra were normalized using a semiautomatic polynomial fitting routine provided by \cite{Folsom16}.
        We provide a journal of observations in Table~\ref{tab:log_obs}.

        \begin{table}[ht]
            \centering
                \begin{tabular}{l l l l l}
                    \hline
                    Date & HJD & S/N & S/N$_{LSD}$ & \phirot \\
                    (2017) & (2,450,000+) &  &  \\
                    \hline
                    28 Oct & 8054.90048 & 159 & 7059 & 0.732 \\
                    29 Oct & 8055.89091 & 178 & 7900 & 0.957 \\
                    30 Oct & 8057.00939 & 147 & 6417 & 1.213 \\ 
                    31 Oct & 8057.99515 & 138 & 5805 & 1.438 \\
                    01 Nov & 8059.05448 & 161 & 7272 & 1.679 \\
                    03 Nov & 8060.95421 & 182 & 8303 & 2.112 \\
                    04 Nov & 8061.95085 & 191 & 8796 & 2.339 \\
                    05 Nov & 8063.08059 & 163 & 7169 & 2.597 \\
                    07 Nov & 8064.99530 & 177 & 7982 & 3.034 \\
                    08 Nov & 8065.99958 & 162 & 7257 & 3.263 \\
                    09 Nov & 8066.91484 & 174 & 7658 & 3.471 \\
                    \hline
                \end{tabular}
            \caption{Journal of ESPaDOnS observations. The columns list the date of observation, the Heliocentric Julian Date, the signal-to-noise ratio by spectral resolution element at 731 nm, the effective signal-to-noise ratio of the Stokes V LSD profiles, and the rotational phase.}
            \label{tab:log_obs}
            \end{table}

    
\section{Results}
\label{sec:results}

    We analyze in this section the photometric and spectroscopic properties of the system, deriving its stellar parameters, and we investigate photometric and line profile variability. Eventually, we derive a brightness map and a magnetic map of the stellar surface from Doppler and Zeeman-Doppler imaging. 

    \subsection{Stellar parameters}
    \label{subsec:stellarparam}
        To derive the stellar parameters, we fitted ESPaDOnS high-resolution spectra with synthetic ZEEMAN spectra \citep{Landstreet88, Wade01, Folsom12} using a $\chi^2$ minimization method based on a Levenberg Marquart algorithm (LMA). The synthetic ZEEMAN spectra are based on MARCS atmosphere models \citep{Gustafsson08} and the VALD line list database \citep{Ryabchikova15}.
        
\begin{figure}
            \centering
            \includegraphics[width=.49\textwidth]{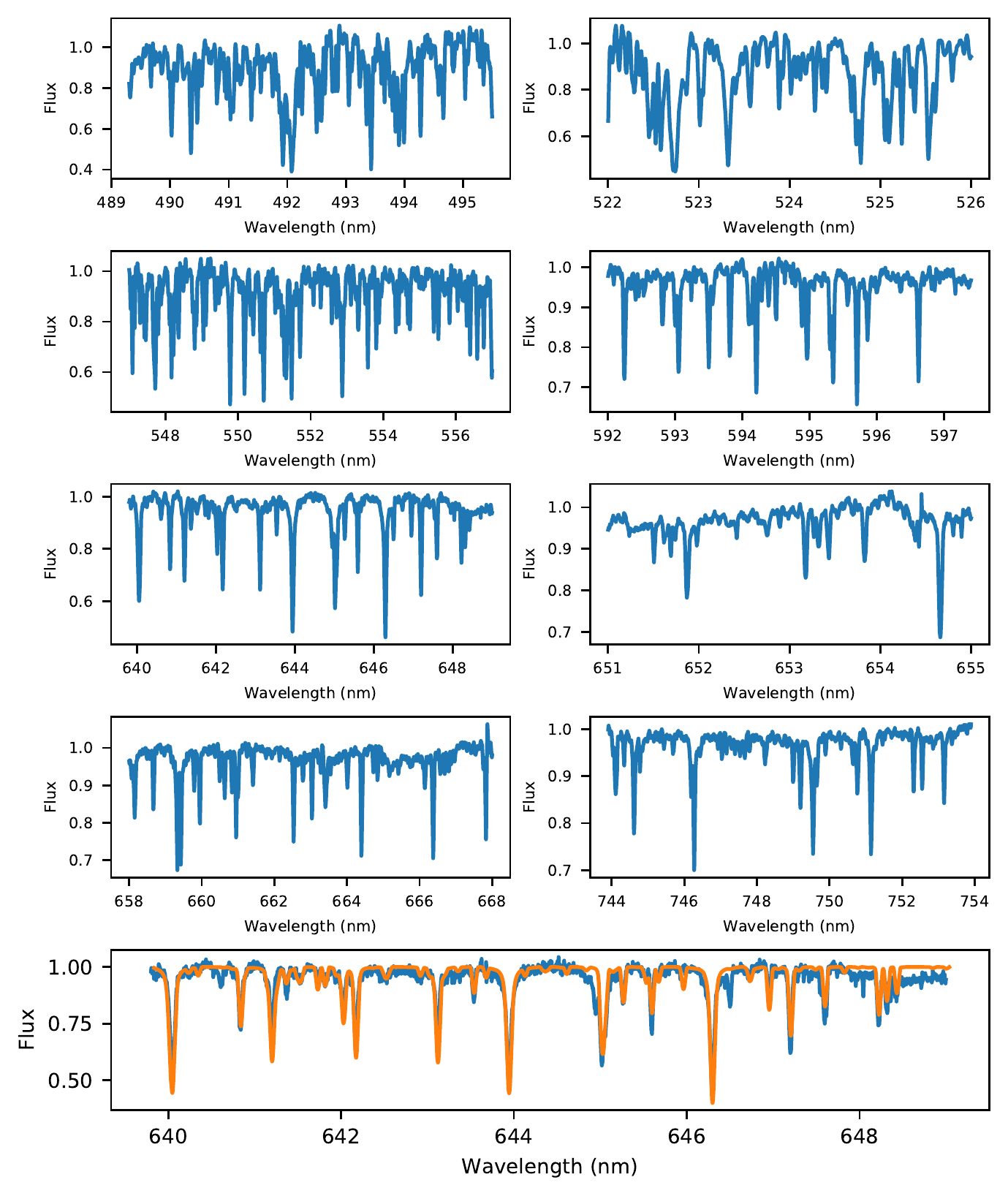}
            \caption{First four rows: Spectral windows used for parameter fitting. Last row: V807 Tau's 639.8 - 649.0 nm spectral window (blue) overplotted with corresponding ZEEMAN synthetic spectrum (orange)}
            \label{fig:lma}
        \end{figure}

        We extracted the effective temperature (\teff), and radial (\vrad), rotational (\vsini), and microturbulent (\vmic) velocities by running the fitting procedure over 8 wavelength windows ranging from 522 to 754 nm, and excluding the regions with tellurics, emission or molecular lines. These are shown in Figure \ref{fig:lma}.
        
        In order to derive those parameters, we set the macro-turbulent velocity (\vmac) to 2.0 \kms, the surface gravity \logg\  to 4.0, and assumed a solar metallicity, which are typical parameters for low-mass TTSs, and used the veiling value derived in Section \ref{subsec:resspectro}.
        We averaged the results from all spectra and wavelength windows, excluding the windows yielding larger values than 2-sigma from the mean, and derived 
         \teff\,=\,3932\,$\pm$\,158\;K, \vsini\,=\,13.3\,$\pm$\,1.8\;\kms, \vrad\,=\,15.9\,$\pm$\,0.3\;\kms\ and \vmic\,=\,2.1\,$\pm$\,0.5\;\kms.
        To lower the uncertainty on the \teff\;estimate, we fitted a mean spectrum by fixing all other parameters to the values found previously and focused on photospheric lines sensitive to \teff\;variation, thus reducing the wavelength range between 550 and 650 nm.
        The resulting value, \teff\,=\,4067\,$\pm$\,59\;K is consistent with the previous estimate and has a much lower uncertainty.
        Using a conversion table from  \cite{Herczeg14}, this \teff\;corresponds to a K6.5-K7 spectral type, in agreement with \cite{Schaefer12}.

        We next derived V807 Tau's luminosity to deduce its mass, radius, age, and internal structure from its location in the Hertzsprung-Russell diagram (HRD).
        We used the V and J band magnitudes derived by \cite{Schaefer12} for the primary from the flux ratio between the three components of the system computed by the authors, combined with its Johnson V \citep{Kharchenko09} and 2MASS J band integrated magnitudes. This yields $({V-J})$\,=\,2.694\,$\pm$\,0.139\;mag and, 
        using intrinsic colors from \cite{Pecaut13} and the interstellar reddening law from \cite{Cardelli89}, we derived $A_{V}$\,=\,-0.11\,$\pm$\,0.19\;mag and $A_{J}$\,=\,-0.03\,$\pm$\,0.05\;mag. We thus assumed that the system suffers negligible extinction on the line of sight. 
        Finally, we used the Gaia DR2 parallax\footnote{At the time of this analysis, Gaia EDR3 had not been released. However, because of the bad RUWE of V807 Tau, none of the two solutions is better than the other one; we thus kept the DR2 value. As information, using the Gaia EDR3 parallax $\pi$ = 5.43 $\pm$ 0.77 mas, we got L = 1.99 $\pm$ 0.71 L$_{\odot}$, R = 2.85 $\pm$ 0.60 R$_{\odot}$, M = 0.596$^{+0.059}_{-0.049}$ M$_{\odot}$, M$_{rad}$ = 0 $\pm$ 0 M$_{\star}$, and R$_{rad}$ = 0 $\pm$ 0 R$_{\star}$.} ($\pi$ = 8.83 $\pm$ 0.66 mas, \cite{Gaia18}) to compute the absolute magnitude in the J band, M$_J$ = 3.42 $\pm$ 0.19, from which we deduced the bolometric magnitude, M$_{bol}$ = 5.05 $\pm$ 0.24, by applying the bolometric correction from \cite{Pecaut13} (BC$_J$ = 1.34 $\pm$ 0.055). 
        Then we use M$_{bol}$ to obtain \lstar = 0.75 $\pm$ 0.17 \lsun\ and the Stefan-Boltzmann law to deduce \rstar = 1.75 $\pm$ 0.18 R$_{\odot}$.
        These estimates are consistent with those reported by \cite{Schaefer12}, \lstar = 0.68 $\pm$ 0.13 \lsun, and by \cite{Akeson19}, \lstar = 0.96$^{+0.17}_{-0.02}$ \lsun, but not with the value derived by  \cite{Herczeg14}, \lstar = 1.82 \lsun, who adopted A$_V$ = 0.5 mag. Combining the stellar radius, the rotational period of the system, and its \vsini, we derive an inclination of i = 41 $\pm$ 10\deg. 
        \begin{figure}
            \centering
            \includegraphics[width=.49\textwidth]{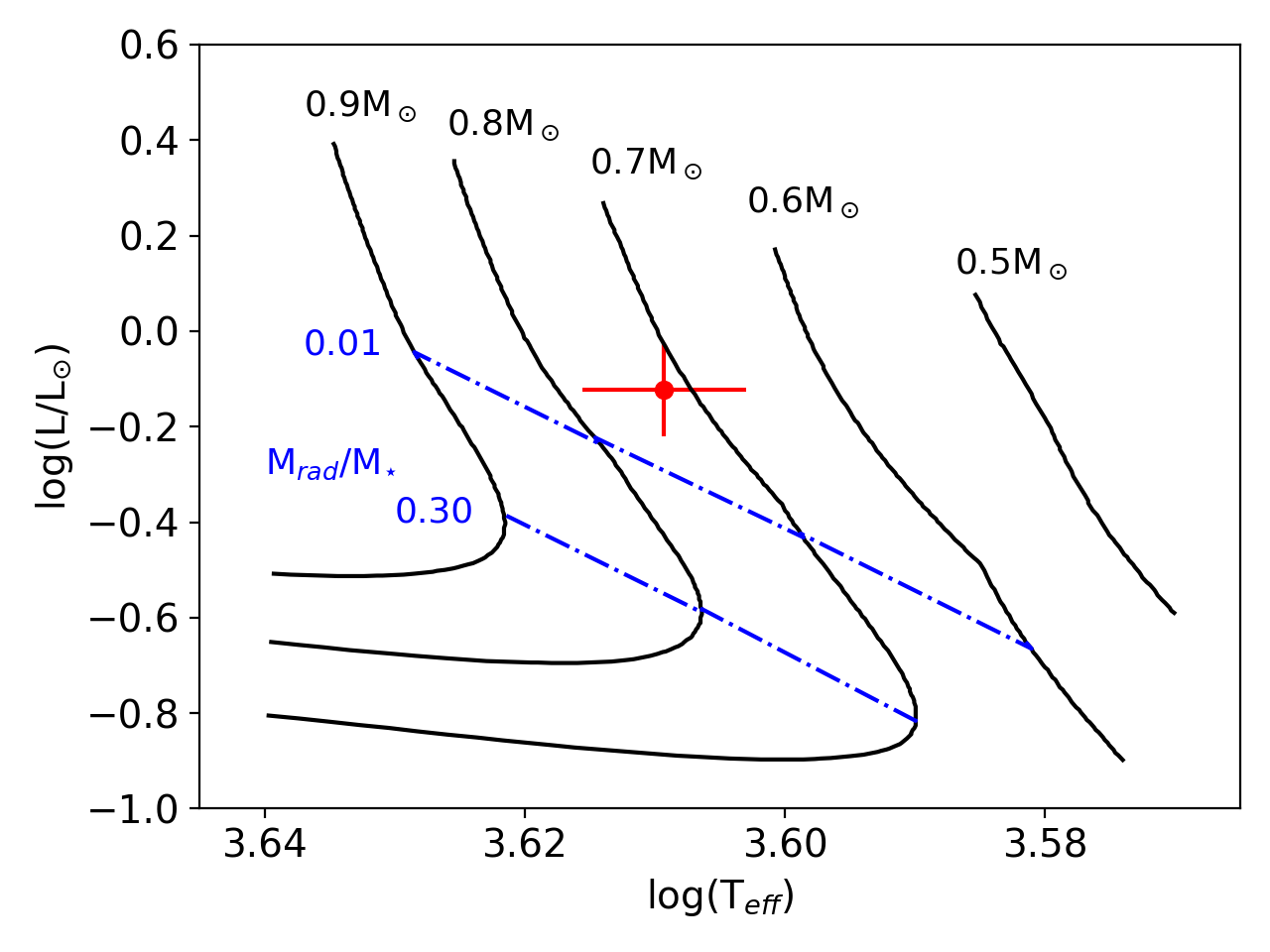}
            \caption{V807 Tau’s position in the HR diagram. The red cross illustrates V807 Tau’s position with corresponding uncertainties. The black curves are the evolutionary tracks from \cite{Baraffe98} PMS models, with the corresponding mass indicated on the top. The blue curves show the position where the radiative core of the star reaches 30\% and 1\% of the stellar mass.}
            \label{fig:hrd}
        \end{figure}

        Using the estimates above, we plotted V807 Tau in the HR diagram, as shown in Figure \ref{fig:hrd}, and used a 2D interpolation routine on \cite{Baraffe98}  pre-main-sequence (PMS) evolutionary tracks to derive M$_\star$ = 0.72$^{+0.07}_{-0.6}$ \msun\ and an age of $\sim$ 1.8 Myr.
        The models also yield the internal structure of the star, with the fractional mass and radius of the radiative core, M$_{rad}$/M$_{\star}$ = 0.0$^{+0.004}_{-0.0}$ and R$_{rad}$/\rstar = 0.0$^{+0.1}_{-0.0}$. V807 Tau  is thus fully or mostly convective.
        We computed the stellar parameters from the CESTAM models as well \citep{Marques13,Villebrun19}, and obtained consistent results, namely M$_\star$ = 0.67$^{+0.04}_{-0.06}$ \msun\ and an age of about 1.7 Myr. 
        
        The stellar parameters of V807 Tau are summarized in Table~\ref{tab:param}. 
        
        \begin{table}
            \centering
            \begin{tabular}{l l}
                \hline
                \hline
                \teff & 4067 $\pm$ 59 K \\
                \vsini & 13.3 $\pm$ 1.8 \kms \\
                $i$ & 53 $\pm$ 5$^{\circ}$ (Sect.~\ref{subsec:resspectropola})\\
                \vrad & 16.99 $\pm$ 0.03 \kms\ (Sect.~\ref{subsec:radvel}) \\
                \lstar & 0.75 $\pm$ 0.17 \lstar \\
                \rstar & 1.75 $\pm$ 0.18 \rsun \\
                \mstar & 0.72 $^{+0.07}_{-0.06}$ \msun \\
                Age & 1.8 Myr \\
                R$_{\rm rad}$ & 0.0$^{+0.1}_{-0.0}$ \rstar \\
                M$_{\rm rad}$ & 0.0$^{+0.004}_{-0.0}$ \mstar \\
                \prot & 4.386 $\pm$ 0.005 d (Sect.~\ref{subsec:resphoto})\\
                \macc & 7 $\pm$ 2 10$^{-10}$ \msunyr\ (Sect.~\ref{subsec:macc}) \\
                
            \hline
            \end{tabular}
            \caption{V807 Tau stellar parameters. The section where the parameter is derived is indicated in the case if it is not in Sect.~\ref{subsec:stellarparam}.}
            \label{tab:param}
        \end{table}
   
    \subsection{Photometric analysis}
    \label{subsec:resphoto}
\begin{figure}
           \centering
           \includegraphics[width=0.45\textwidth]{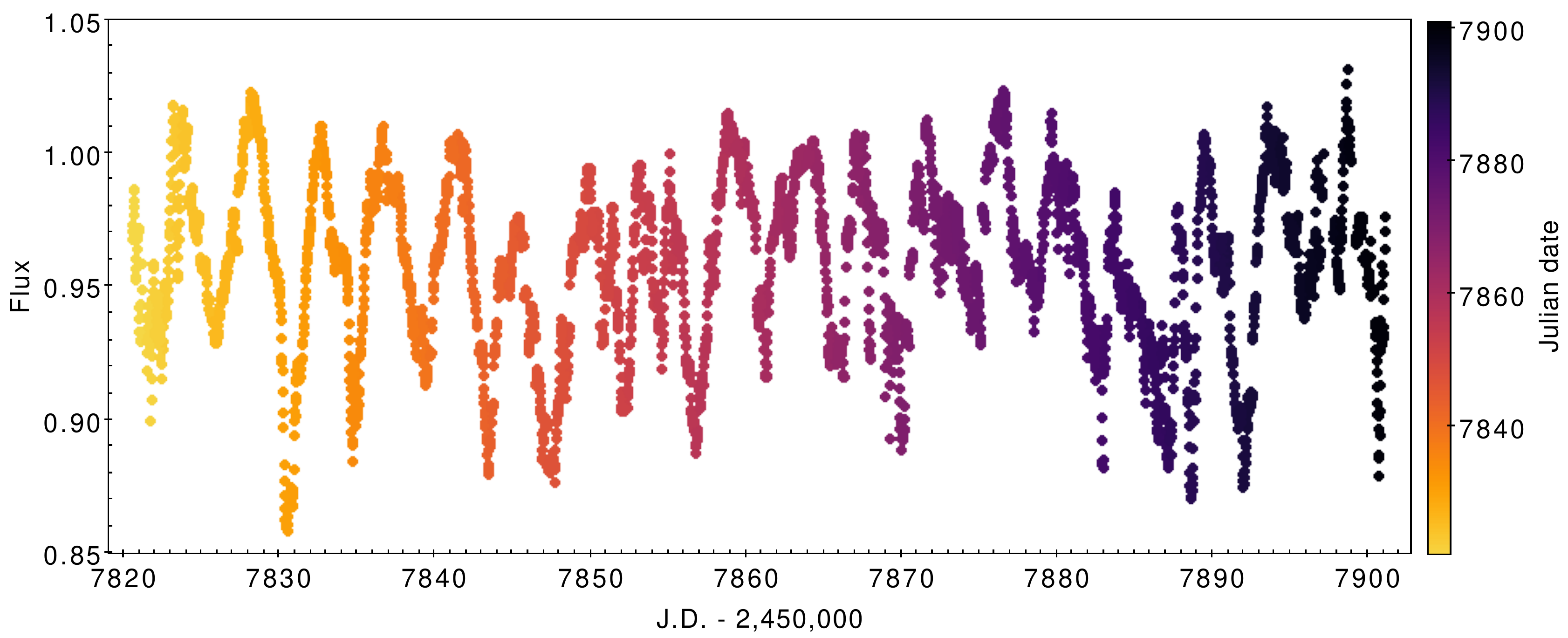}
           \includegraphics[width=0.45\textwidth]{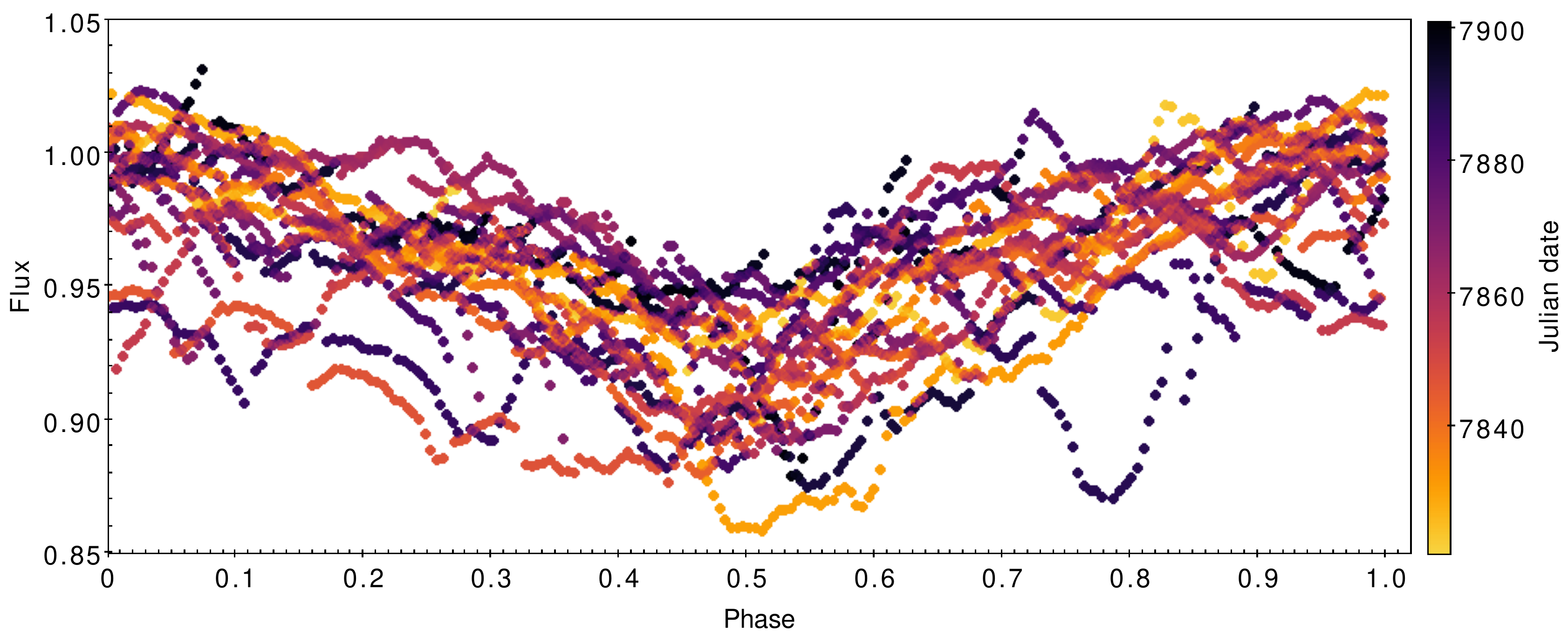}
           \caption{{\it Top:} Kepler K2 light curve of V807 Tau.
           {\it Bottom:} Light curve folded in phase using a 4.38 d-period. The origin of time is arbitrarily set as JD 2,457,819.55, in order to locate the photometric minimum at phase 0.5. The color code scales with the Julian Date.}
           \label{fig:k2lc}
       \end{figure}

    A \textit{K2} light curve, shown in Fig.~\ref{fig:k2lc}, was obtained for V807 Tau over 80 days in Spring 2017.
    It exhibits recurrent dips that appear periodic. 
    A CLEAN periodogram analysis \citep{Roberts87} yields \prot = 4.386 $\pm$ 0.005~d, which is consistent the \prot = 4.3784 $\pm$ 0.002~d period derived by \cite{Rebull20} from the same data set. 
    The K2 light curve folded in phase is shown in Fig.~\ref{fig:k2lc}. 
    It confirms the periodicity of the photometric variations, even though there are significant residual variations beyond the periodic component. 
    The light curve itself was classified as a dipper by \cite{Rebull20}, a class that often shows variable dips from one rotational cycle to the next.
    A more detailed discussion of V807 Tau's K2 light curve is provided in Appendix~\ref{ap:v807lc}.
    
    We assume here that the photometric period measures the stellar rotational period, and we thus adopt the following ephemeris throughout the paper to compute the rotational phase: 
       \begin{equation}
           JD (d) = 2,457,819.55  + 4.38E,
           \label{eq:ephemeris}
       \end{equation}
       where E is the rotational cycle.

    The $\sim$0.005~d uncertainty attached to the period determination does not significantly impact the relative rotational phase within the series of ESPaDOnS spectra: Over a timescale of 12 days, the differential phase shift between the first and the last spectra would only amount to d\phirot = $\delta$t dP/P$^2$ = 0.005*12/4.38$^2$ = 0.003, which is negligible indeed. 
    Even over the much longer timescale elapsed between the K2 light curve and the spectral data set, namely about 150 days, the phase shift due to the period uncertainty amounts to only d\phirot=0.04. 
    We can thus confidently use the K2 light curve and its ephemeris to estimate the rotational phase at which the ESPaDOnS spectra were obtained.

\begin{figure}
           \centering
           \includegraphics[width=0.45\textwidth]{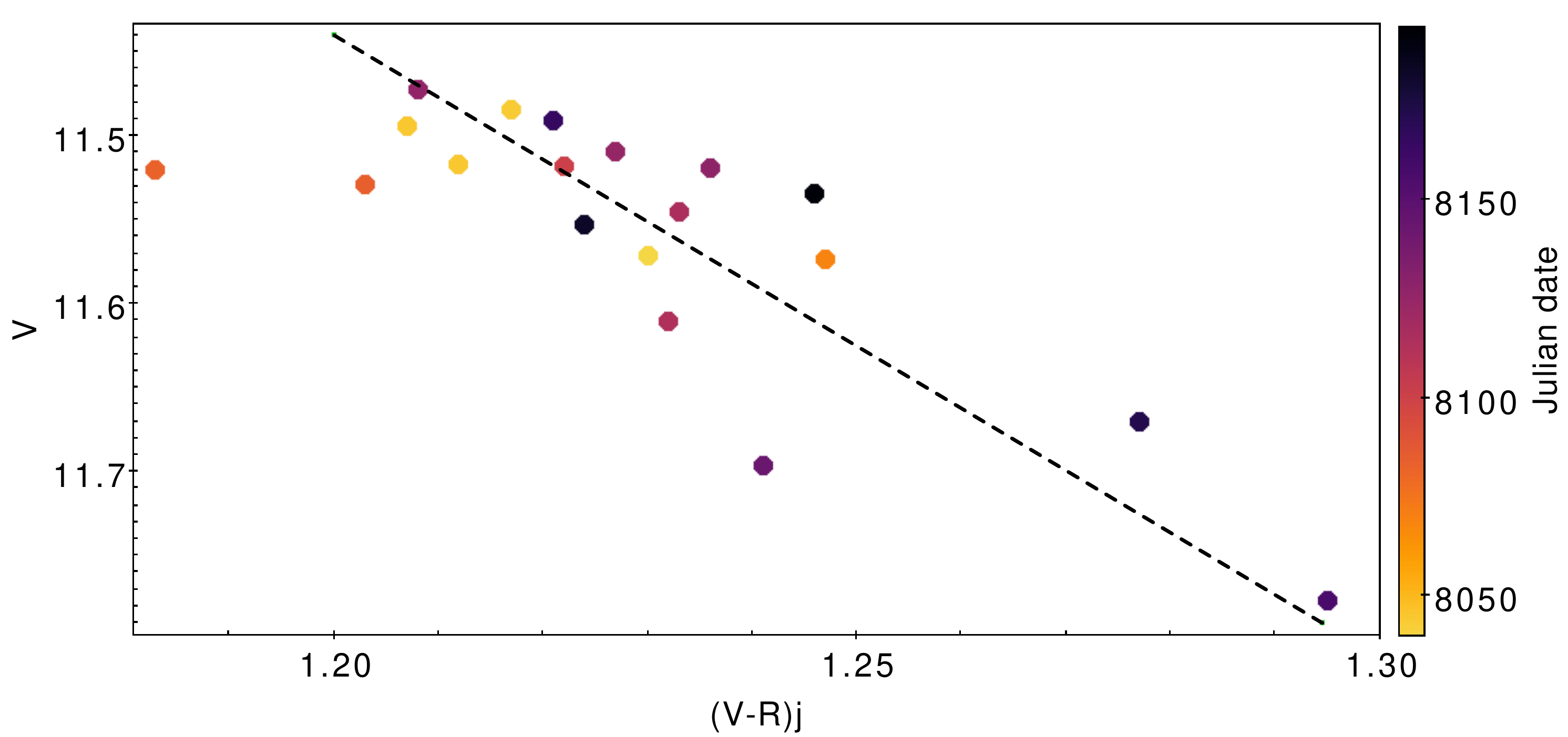}
           \includegraphics[width=0.45\textwidth]{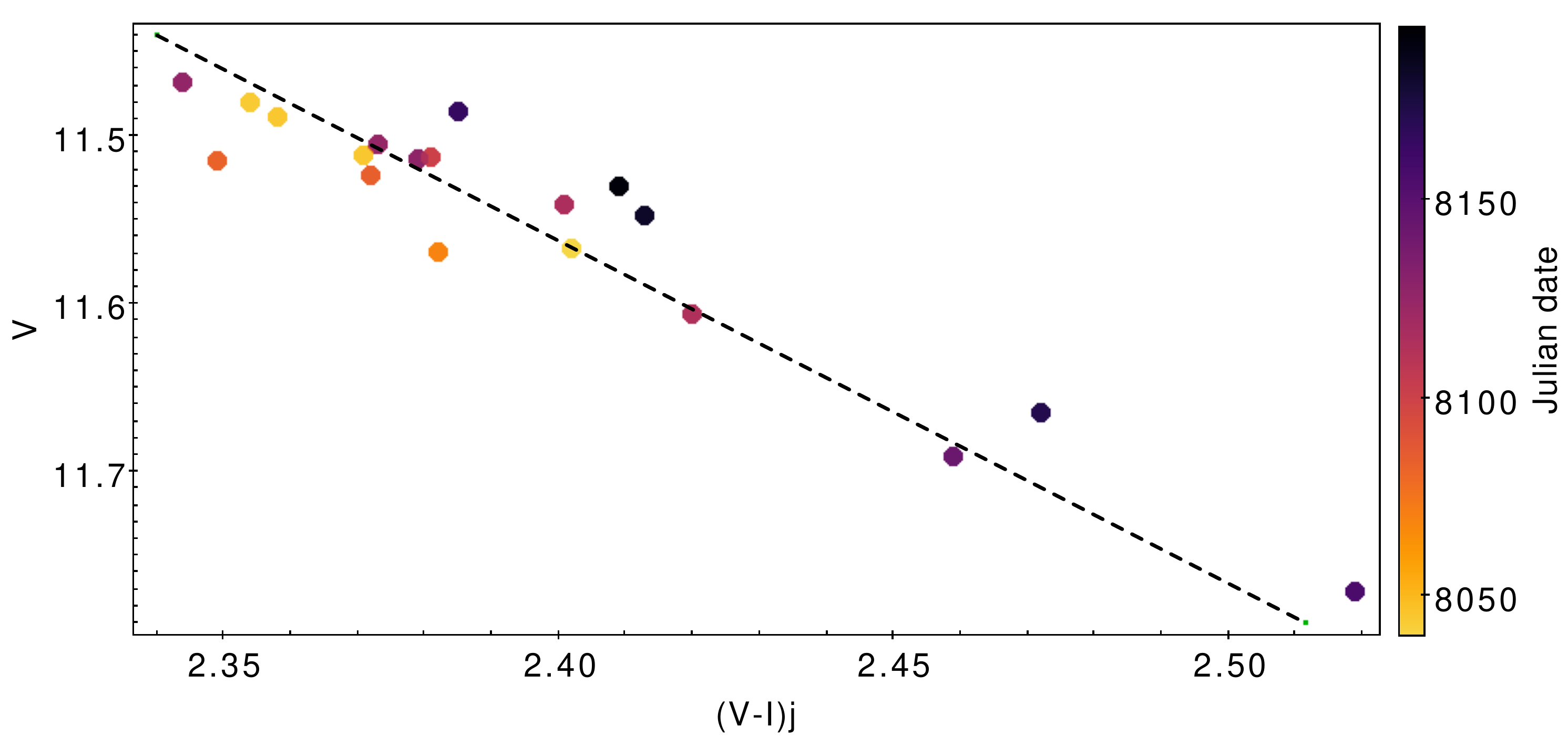}
           \caption{V-(V-R)$_J$ ({\it top}) and V-(V-I)$_J$ ({\it bottom}) color-magnitude diagrams from CrAO multicolor photometry. The rms uncertainty is 0.013 mag in V and 0.020 mag in colors.} The color code scales with the Julian Date.  The dashed line depicts the color slope of interstellar extinction.  The system becomes redder when fainter, following a slope consistent with ISM reddening.
           \label{fig:craocolmag}
       \end{figure}

    The loose sampling rate of the CrAO photometry allows us to recover neither the periodicity in the photometric variations nor the epoch of the photometric minimum. 
    For these two quantities we therefore fully rely on the previous Kepler K2 light curve. 
    CrAO photometry nevertheless provides color information.      
    Color-magnitude diagrams are shown in Fig.~\ref{fig:craocolmag}. 
    The system becomes redder when fainter, and the color changes are consistent with reddening by ISM-like grains \citep{Cardelli89}. 
    While this color behavior supports the dipper light curve classification of this source, with recurrent dimming of the central star being due to circumstellar dust passing onto the line of sight, we cannot exclude the contribution from cold stellar spots as their color slope has been shown to be in the same range as that of circumstellar dust \citep{Venuti15}.

    \subsection{Spectroscopic analysis}
    \label{subsec:resspectro}
    We present here the analysis of the ESPaDOnS spectroscopic time series.
    We first analyze the radial velocity variability of the star and then focus on the circumstellar line variability.
    The major emission lines seen in V807 Tau's spectrum are the Balmer H$\alpha$, H$\beta$, and H$\gamma$ lines, as well as the Ca~II infrared triplet (IRT) and the He~I line 587.6 nm line. In addition, we also analyze the 
 Na I D 588.9 nm doublet and the KI 770 nm line that mainly appear in absorption.
    
    \subsubsection{Radial velocity}
    \label{subsec:radvel}
\begin{figure}
            \centering
            \includegraphics[width=.45\textwidth]{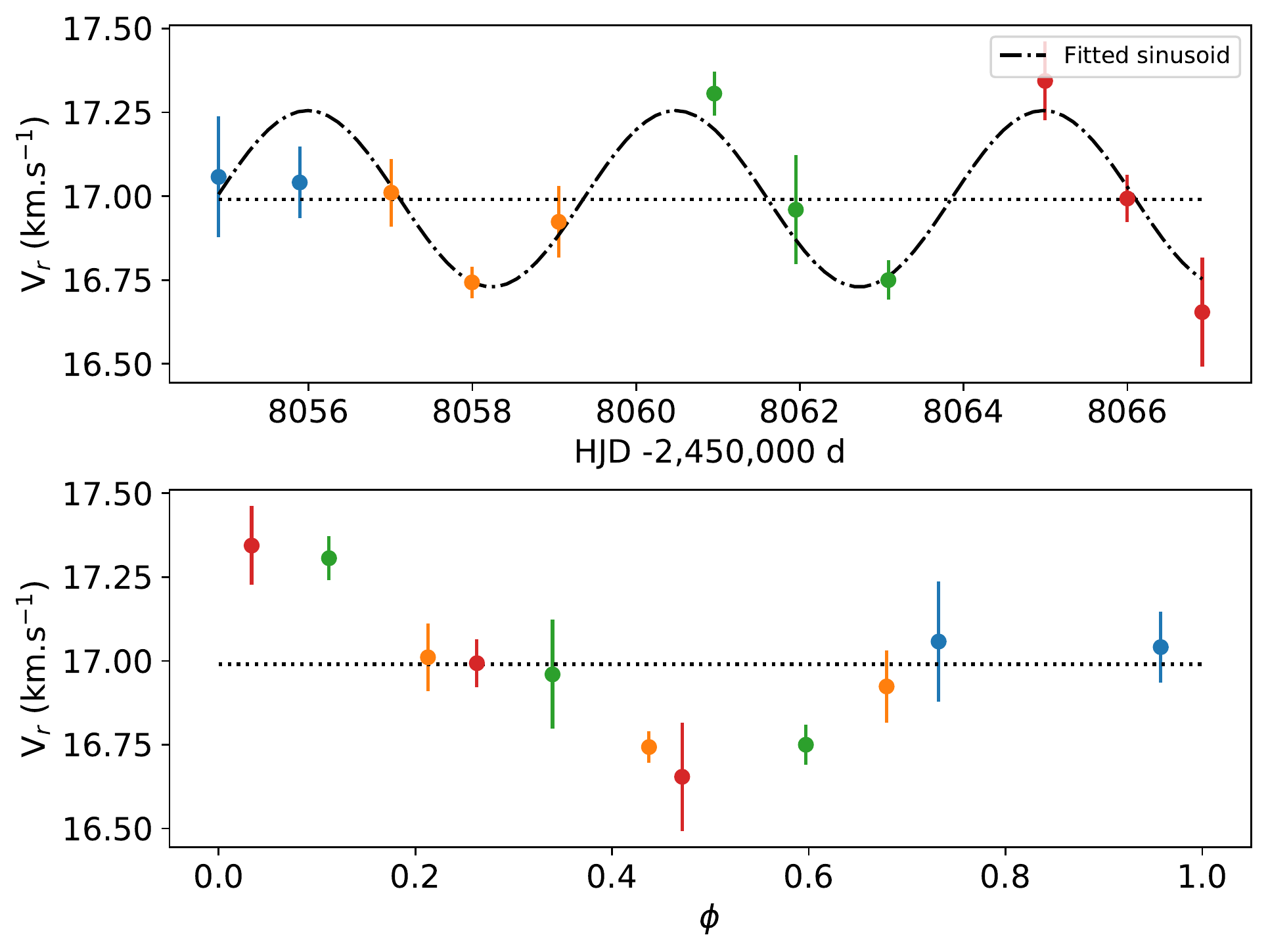}
            \caption{{\it Top:} Radial velocity curve of V807 Tau (dots - each color corresponds to a rotational cycle) fitted with a sinusoid curve at a fixed period of 4.38 d (dash-dotted curve).
            {\it Bottom:} Radial velocity curve folded in phase using a 4.38 day period. The origin of time was set to HJD 2,458,051.69, i.e., 53 periods after the origin of time of the K2 ephemeris (Eq.~\ref{eq:ephemeris}) in order to set the first ESPaDOnS spectrum at cycle 0 while keeping the fractional rotational phase the same. }
            \label{fig:rvcurve}   
        \end{figure}

        To measure V807 Tau's radial velocity (\vrad) variations, we performed a cross-correlation between each spectrum and a photospheric template.
        The method consists of applying a velocity shift between the star and the template, broadened to the same \vsini, over several wavelength windows and computing a linear correlation coefficient.
        This technique applied to a range of velocity shifts yields a curve of correlation coefficient versus velocity shift called the cross correlation function (CCF).
        The location of the peak of the CCF provides an estimate of the \vrad\ difference between the star and the template.
        As a template, we chose V819 Tau, a WTTS with a K7 spectral type, \vrad = 16.6 \kms, and \vsini = 9.5 \kms\ \citep{Donati15}.
        We used 15 different wavelength windows from 467 to 787.7 nm to derive the CCF, in order to obtain a statistical average and rms uncertainty of \vrad.
        We fitted all CCFs by a Gaussian function and we averaged the results over the 15 windows, using the corresponding standard deviation as uncertainty.
        Finally, we excluded the values discrepant by more than 1~$\sigma$ from the analysis, for which the cross correlation failed.
        This removed 3 complete windows (639-640 nm, 651-655 nm, and 738.5-758.5 nm), as well as 45 of the 132 remaining values.
        The results are summarized in Table~\ref{tab:rad_vel} and the radial velocity curve is shown in Fig.~\ref{fig:rvcurve}.
       
        We derive a mean \vrad = 16.99 $\pm$ 0.03 \kms, from a quadratic mean of individual errors, which is consistent with the values found by \cite{Schaefer12} between early 2001  (16.22 $\pm$ 1 \kms) and late 2002 (16.35 $\pm$ 1 \kms). These estimates are consistent within 3$\sigma$ with the ones they measured in late 2003 (15.55 $\pm$ 1 \kms), 2006 (15.40 $\pm$ 1 \kms), and 2010 (14.80 $\pm$ 1 \kms) as well. They suggested that this downward trend is consistent with the orbital motion of the wide A-B pair.

        The radial velocity curve appears to be modulated, and we fitted a sinusoid curve that yielded a period of 4.49 $\pm$ 0.16 d, consistent with the more accurate period derived from the Kepler K2 light curve.
        Furthermore, a Lomb-Scargle periodogram of the radial velocity curve yields a peak at f$\sim$0.22 d$^{-1}$ (i.e., P$\sim$4.55 d) with a FAP$\sim$0.1, consistent with the sinusoid fit and with the stellar rotation period.
        To exclude the hypothesis of a monotonic radial velocity curve due to the error bars, we performed the following statistical test.
        We generated 1,000 radial velocity curves, where each point is randomly set to a value within the error bars shown in Fig.~\ref{fig:rvcurve}, following a Gaussian distribution.
        Then, for each radial velocity curve, we computed the sum of squared error (SSE) defined by:
        \begin{equation}
            SSE = \sum_i (y_i - f(x_i))^2,
        \end{equation}
        where $y_i$ is the data point and $f(x_i)$ the model at the coordinate $x_i$.
        We thus obtained a set of 1,000 SSE derived with a sine curve as a model, and another set with a constant model value at the mean radial velocity. 
        We compared the distributions of the two sets of SSE using an Anderson-Darling test \citep{Anderson52, Anderson54, Scholz87}. 
        The test rejects the null hypothesis of the two distributions being the same with a significance level lower than 0.1\%. This confirms that the radial velocity curve is modulated on the stellar rotation period, as expected from stellar spots \citep{Vogt83}.
         Assuming a dominant single dark spot, it faces the observer when the radial velocity reaches its mean value, going from maximum to minimum.
        This occurs here at $\phi \sim$0.3, which is not consistent with the photometric minimum occuring at $\phi \sim$0.5 in Fig.~\ref{fig:k2lc}. The difference is larger than the phase uncertainty extrapolated from the Kepler K2 light curve to the ESPaDOnS observations. 
        This suggests that the spot inducing the radial velocity variations lies at a different azimuth than the corotating circumstellar structure responsible for the dips seen in the K2 light curve.
        
        

    \subsubsection{Veiling}

        The so-called veiling is the filling in of photospheric lines by the additional continuum flux emitted by the hot spot arising from the accretion shock at the stellar surface. In order to measure veiling in V807 Tau's spectra, we used the same photospheric template and the same wavelength windows as those adopted to derive its radial velocity.
        However, \cite{Rei18} revealed a line-dependent veiling with an additional component seen in the strongest photospheric lines. 
        We thus measured veiling on the weakest lines to reduce this effect.
        The \vsini\ broadened and \vrad\ corrected template spectrum is fitted to V807 Tau's spectrum using an LMA algorithm, adding the veiling as a free parameter with the following equation:
        \begin{equation}
            I(\lambda) = \frac{I_t(\lambda)(1+r)}{1+I_t(\lambda)r},
        \end{equation}
        where $I_t$ is the intensity of the template and $r$ the fractional veiling.
        We thus derived a mean veiling of 0.31 $\pm$ 0.16, indicative of active accretion onto the star, but without evidence for significant variability nor modulation over the timeframe of the observations.
        The uncertainty corresponds to the standard deviation of the values measured on 15 wavelength windows.

    \subsubsection{Balmer lines}
    \label{subsec:balmer}

        We investigate the Balmer emission lines H$\alpha$, H$\beta$, and H$\gamma$ in order to investigate the line shape modulation expected to arise from a non-axisymmetric magnetospheric accretion process.
        We computed residual line profiles using the same photospheric template as before, rotationally broadened and corrected for veiling.

        \begin{figure*}
            \centering
            \begin{subfigure}{.24\textwidth}
                \includegraphics[width=\textwidth]{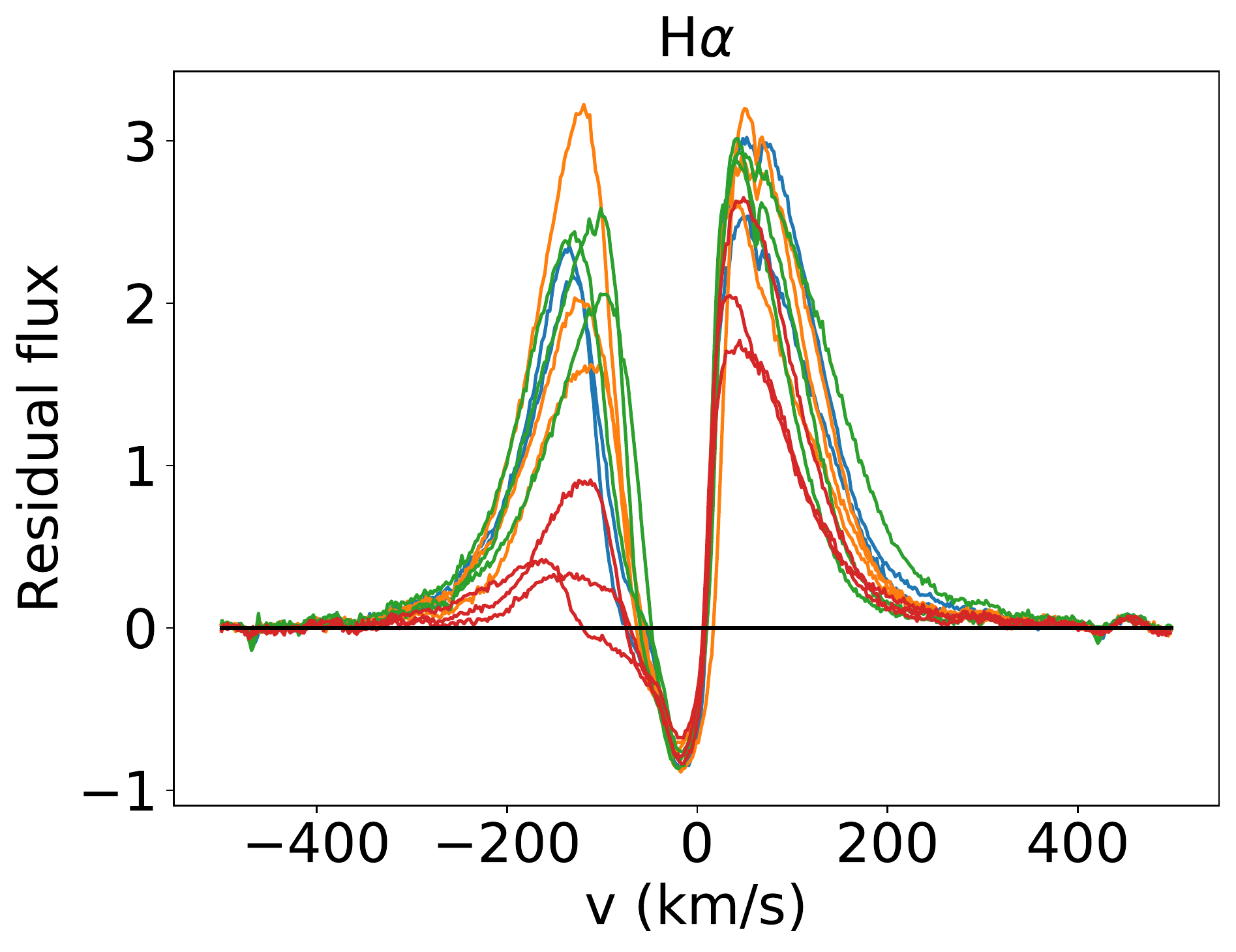}
                \label{fig:resHa}
            \end{subfigure}
            \begin{subfigure}{.24\textwidth}
                \includegraphics[width=\textwidth]{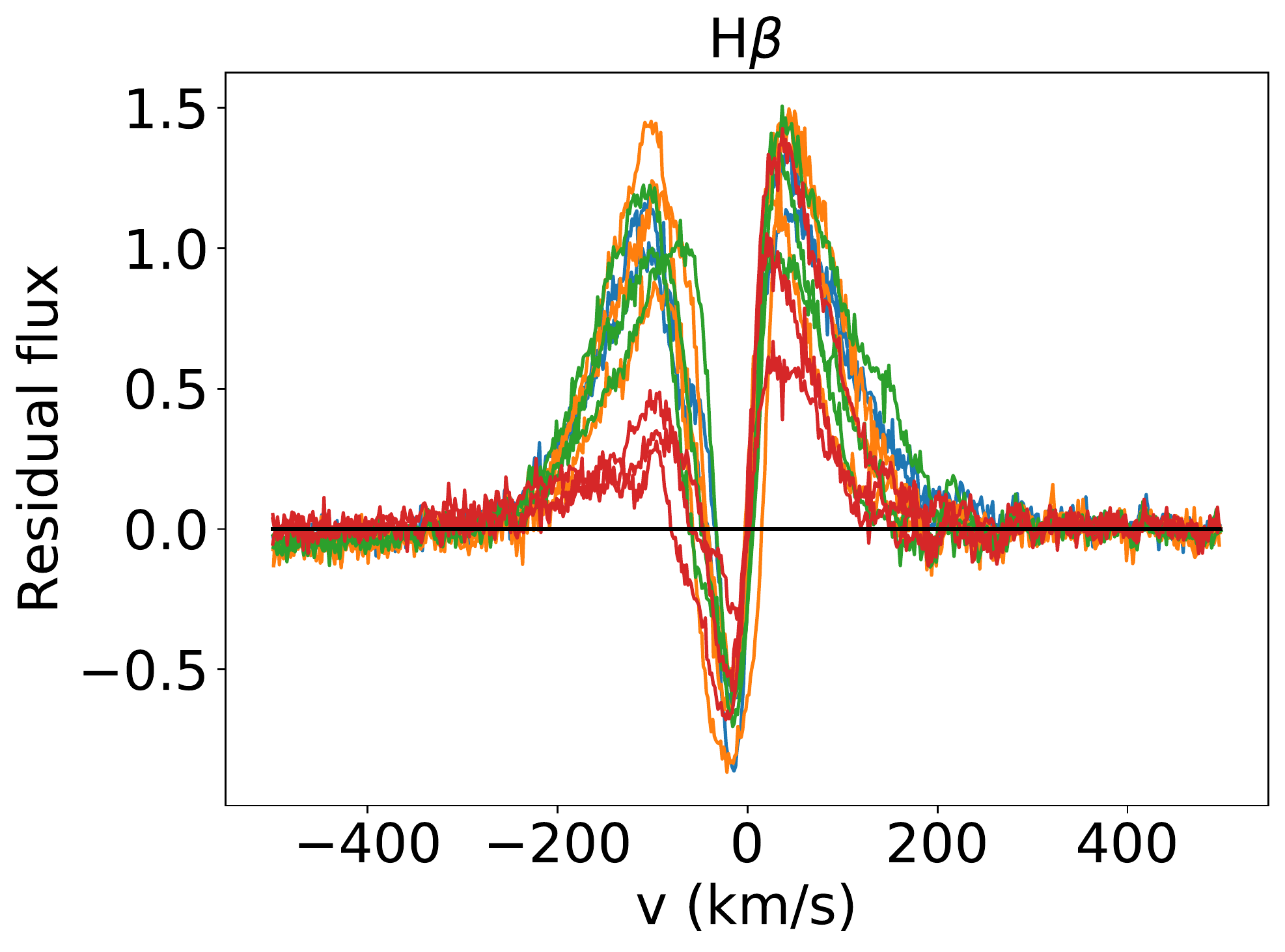}
                \label{fig:resHb}
            \end{subfigure}
            \begin{subfigure}{.24\textwidth}
                \includegraphics[width=\textwidth]{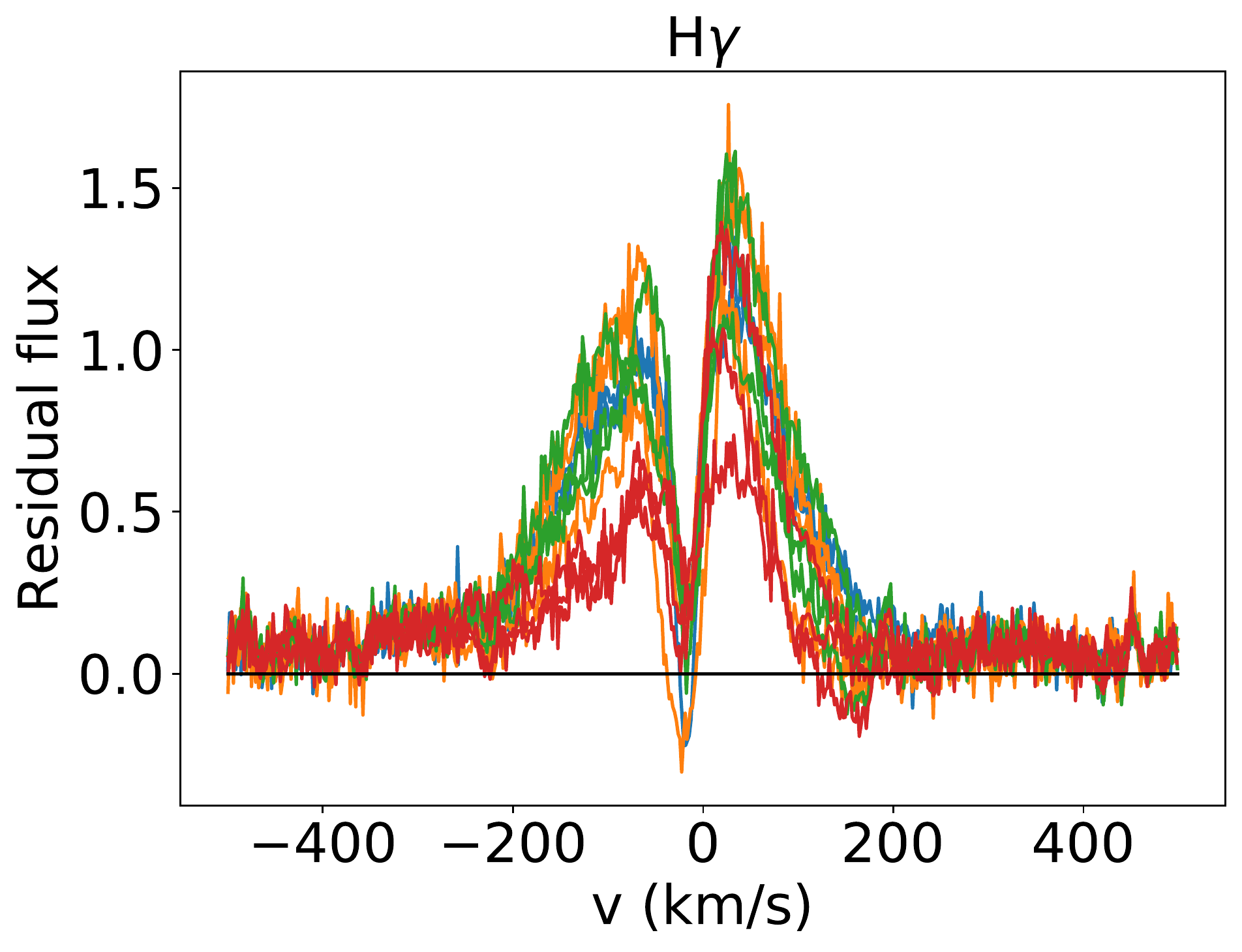}
                \label{fig:resHg}
            \end{subfigure}
            \begin{subfigure}{.24\textwidth}
                \includegraphics[width=\textwidth]{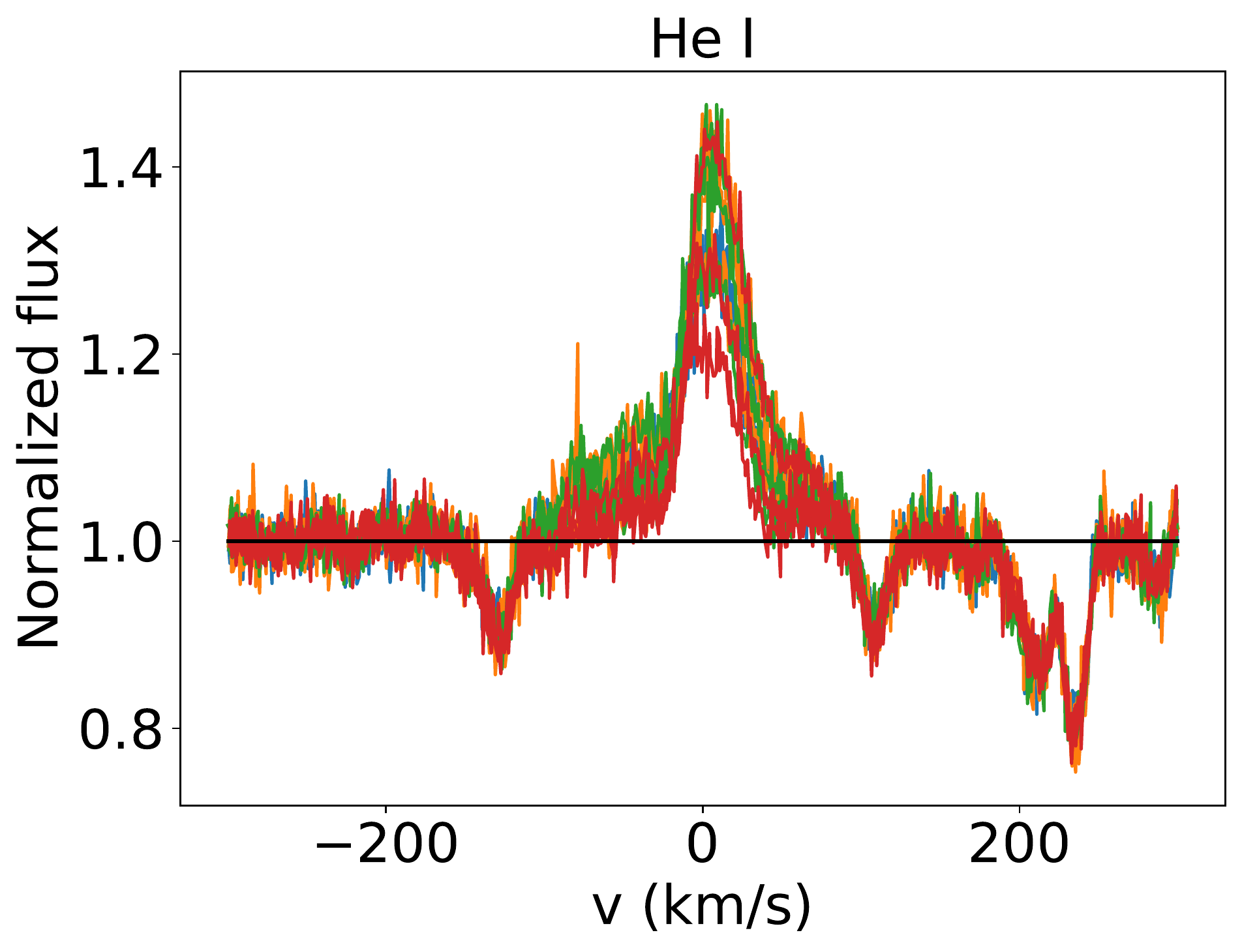}
                \label{fig:HeSup}
            \end{subfigure}
            
            \begin{subfigure}{.24\textwidth}
                \includegraphics[width=\textwidth]{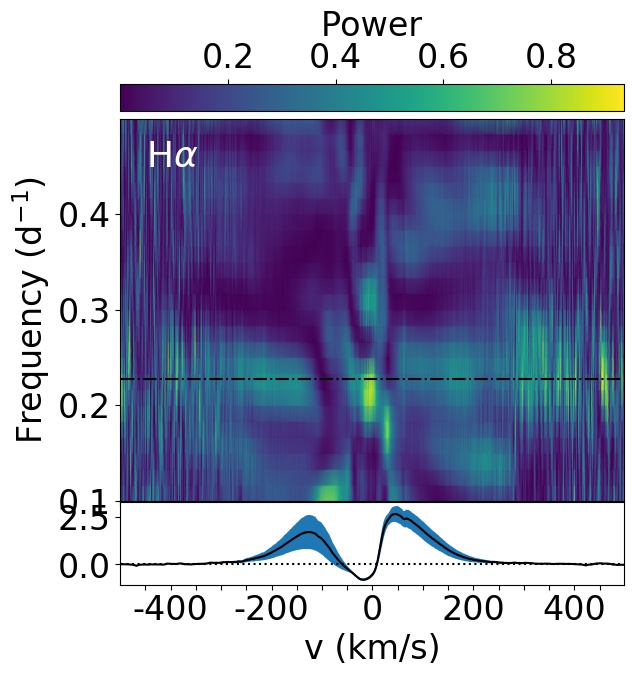}
                \label{fig:periodoHa}
            \end{subfigure}
            \begin{subfigure}{.24\textwidth}
                \includegraphics[width=\textwidth]{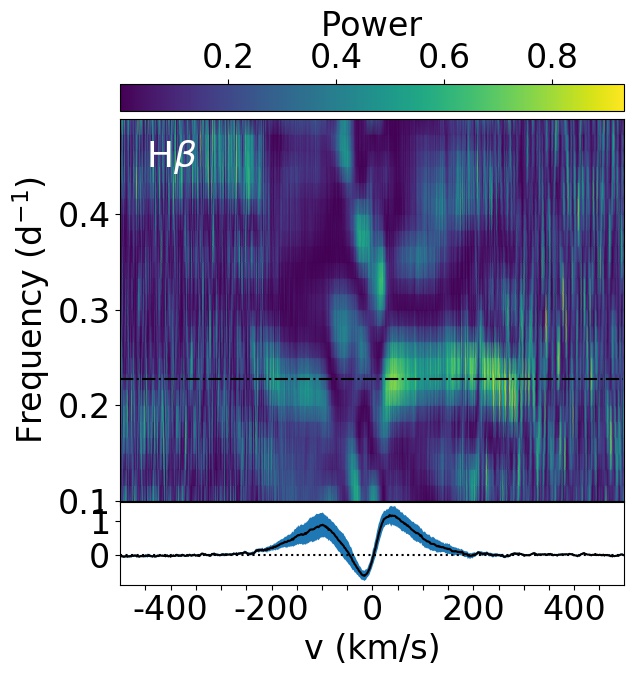}
                \label{fig:periodoHb}
            \end{subfigure}
            \begin{subfigure}{.24\textwidth}
                \includegraphics[width=\textwidth]{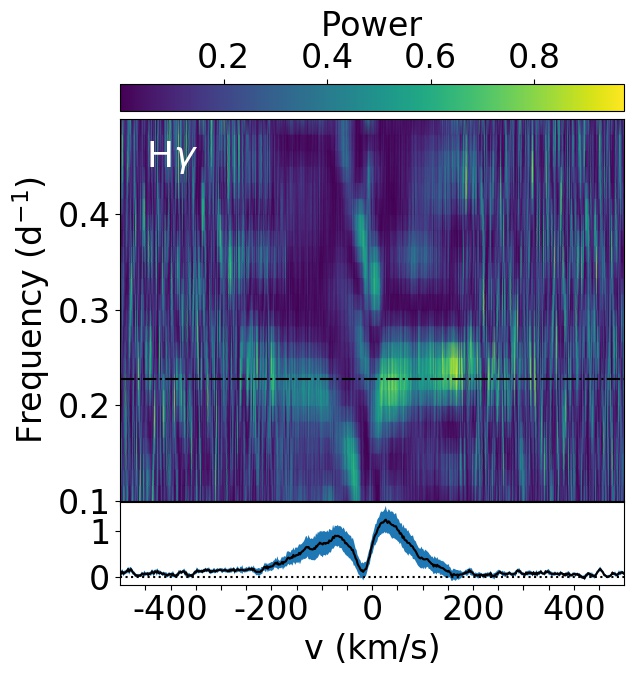}
                \label{fig:periodoHg}
            \end{subfigure}
            \begin{subfigure}{.24\textwidth}
                \includegraphics[width=\textwidth]{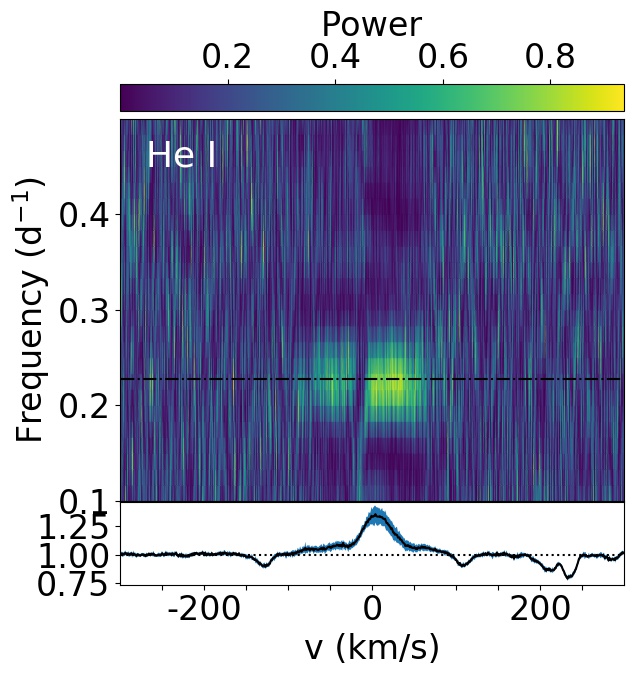}
                \label{fig:He2dperiodo}
            \end{subfigure}
            \caption{{\it Top row:} Residual Balmer line and He~I line profiles. The different colors correspond to successive rotational cycles. {\it Bottom row:} 2D periodograms of residual Balmer line and He~I line profiles. Light yellow represents the highest power of the periodogram. The bottom panel shows the mean residual line profile (black) and its variance (blue). The dashed-dotted line indicates the rotational period of the star.}
            \label{fig:reslines}  
        \end{figure*}

        The Balmer line profiles are shown in Fig.~\ref{fig:reslines}. 
        The profiles are double-peaked with a deep central absorption. They are reminiscent of the Balmer line profiles of AA Tau \citep{Bouvier03, Bouvier07a}, the prototype of dippers,   where the authors interpreted the central component as absorption due to low-velocity material from the funnel flow close to the disk plane.

\begin{figure}
            \centering
            \includegraphics[width=.44\textwidth]{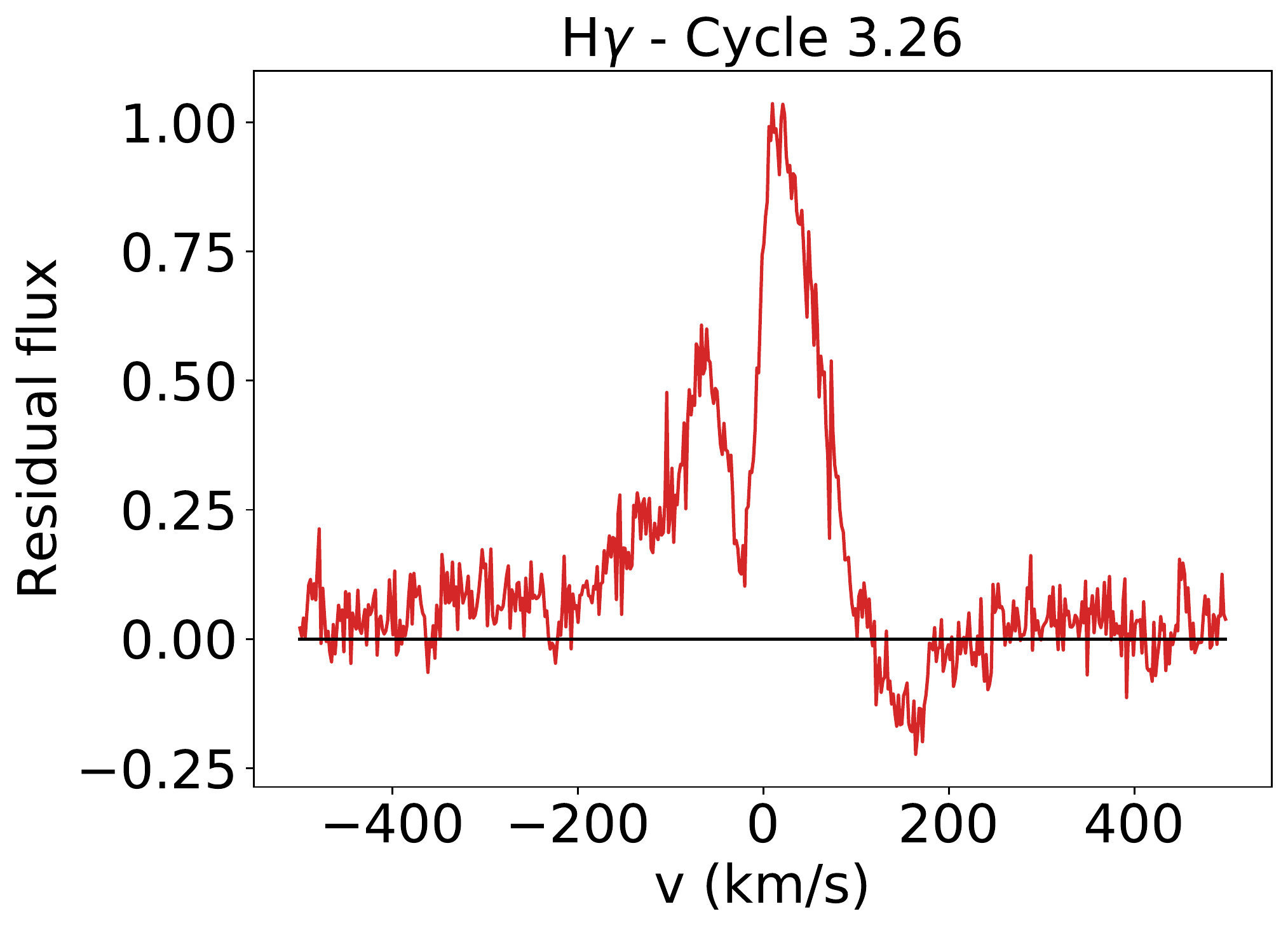}
            
            \includegraphics[width=.43\textwidth]{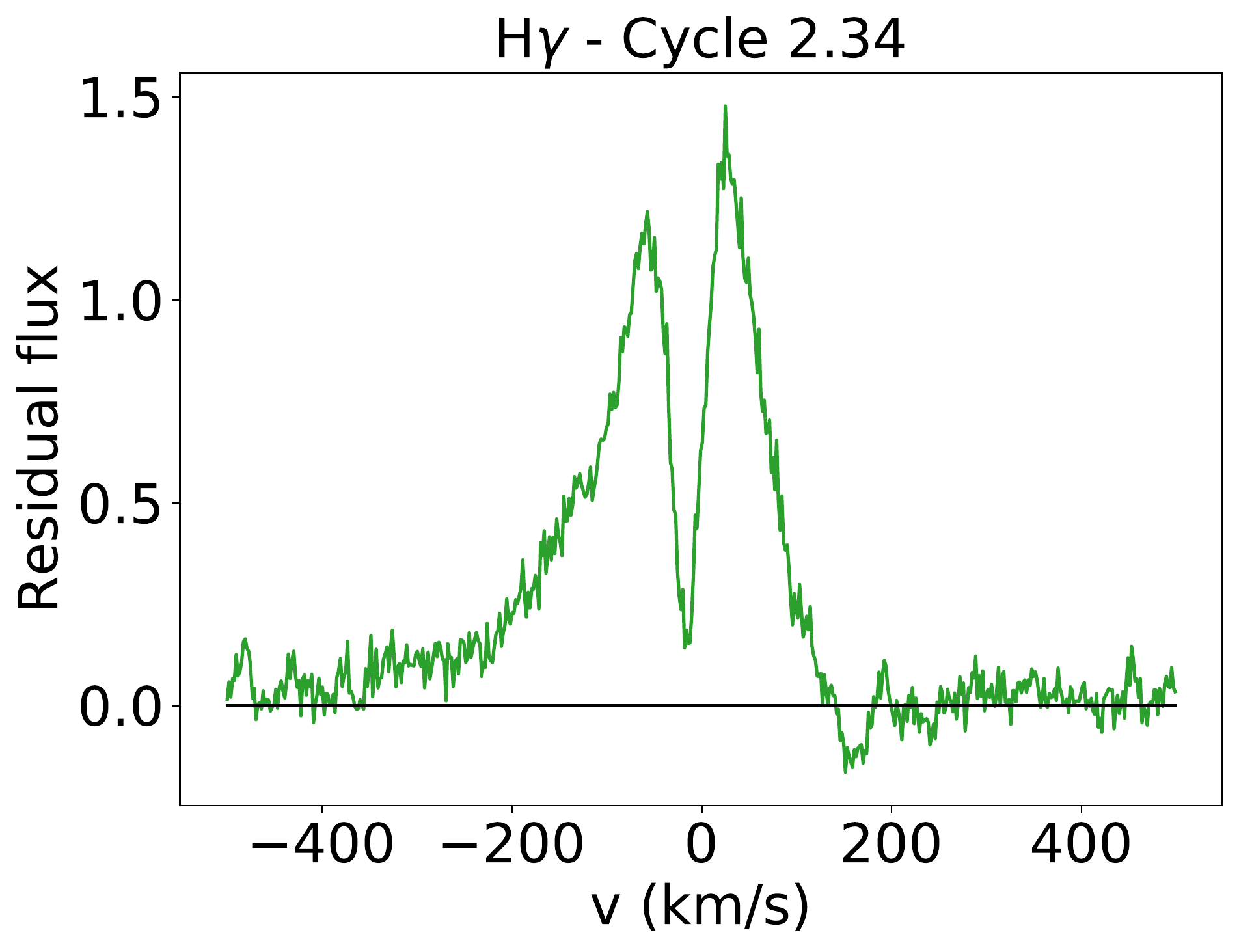}
        
            \caption{H$\gamma$ residual line profiles  exhibiting redshifted absorption components at phase 3.26 \textit{(top)} and 2.34 \textit{(bottom)}.}
            \label{fig:Hgphase}
        \end{figure}

\begin{figure*}
            \centering
            \includegraphics[width=.9\textwidth]{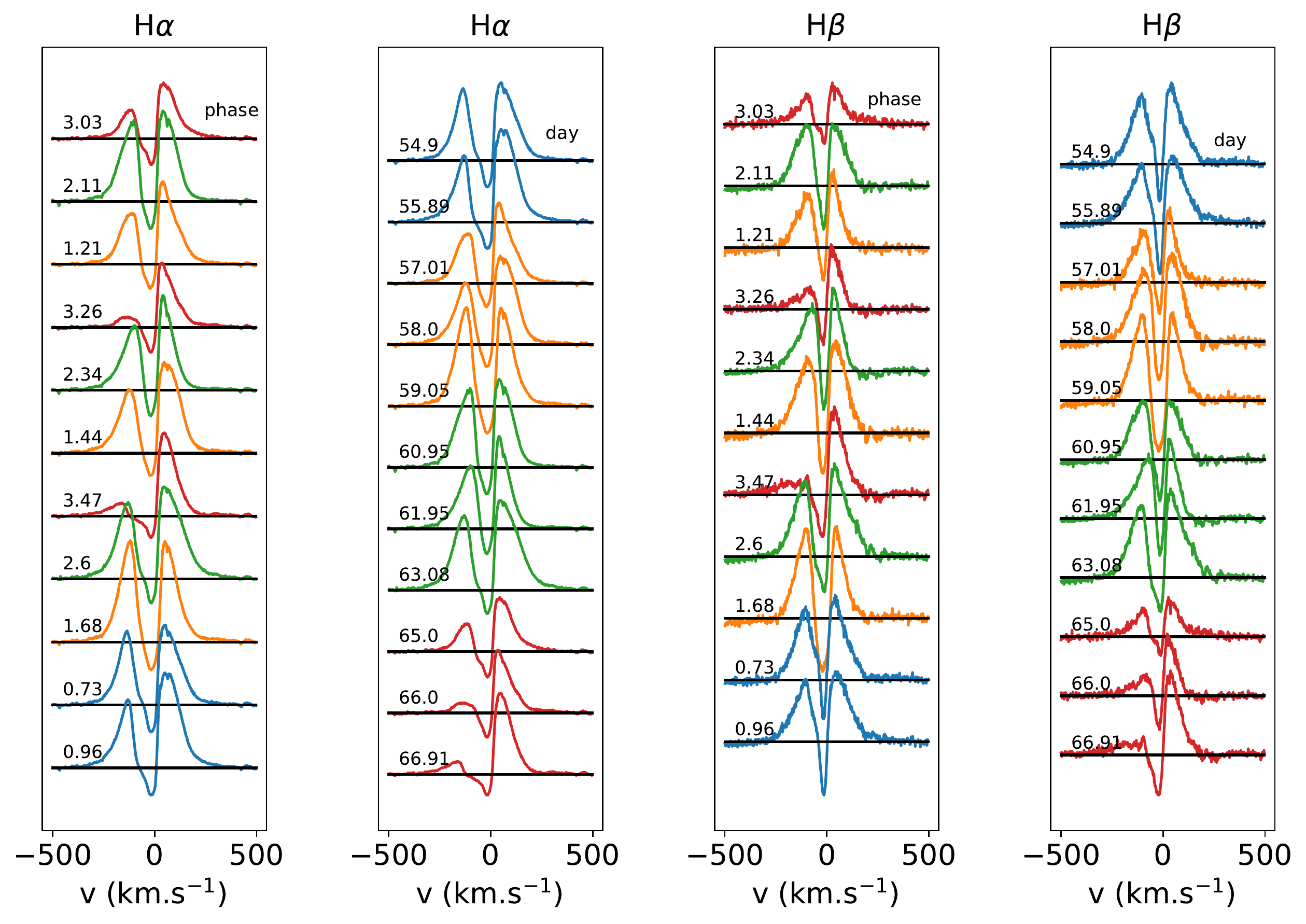}
            \includegraphics[width=.9\textwidth]{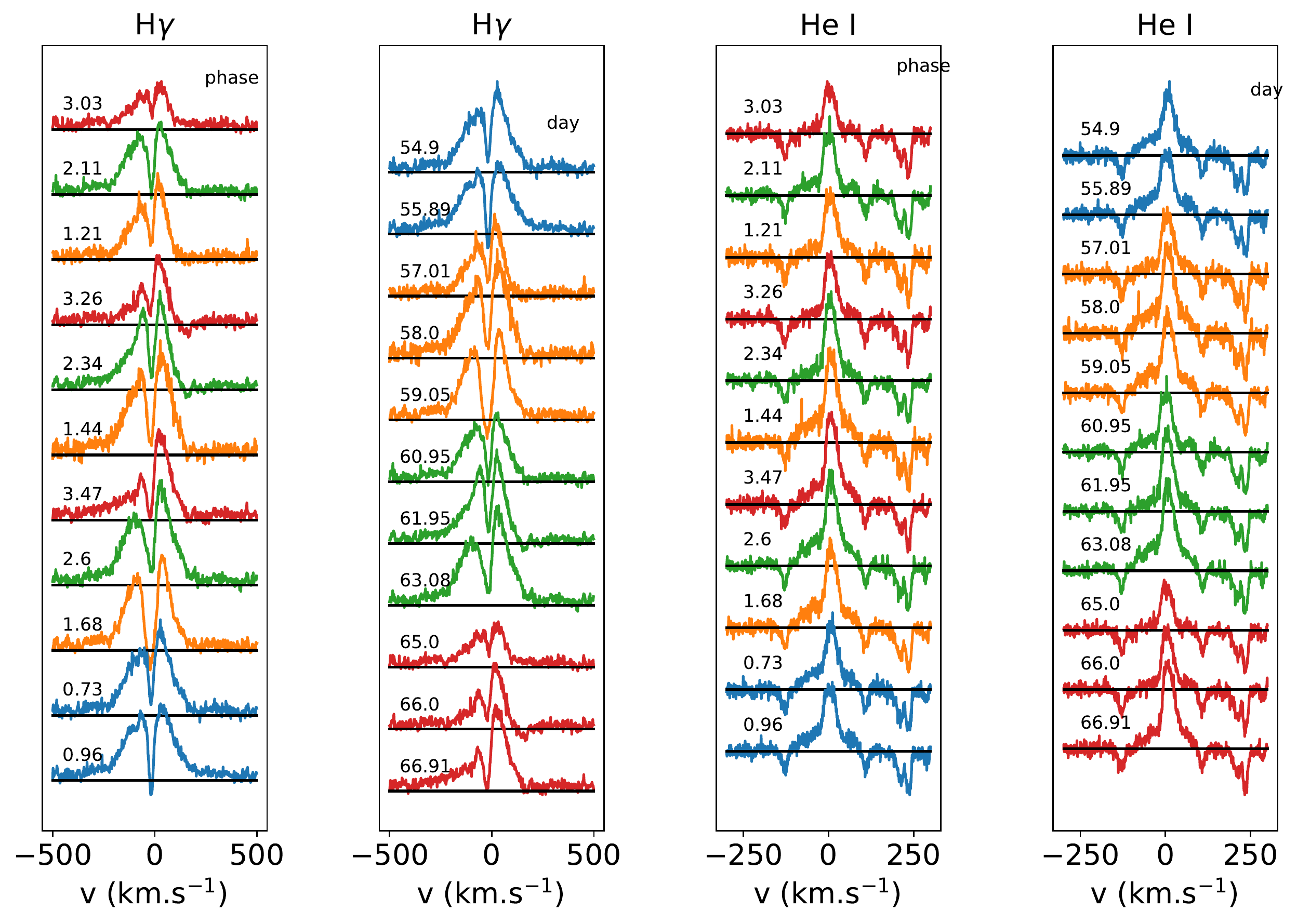}
            \caption{{\it From left to right, top to bottom panels:} Residual profiles of H$\alpha$, H$\beta$, H$\gamma$ lines, and He I profile, sorted by phase and date.}
            \label{fig:columnsprof}
        \end{figure*}

        The profiles are strongly variable in the blue and red emission wings.
        A couple of H$\gamma$ profiles, more clearly shown in Fig.~\ref{fig:Hgphase}, exhibit shallow redshifted absorption components below the continuum up to a velocity of +180 \kms. These inverse P Cygni (IPC) profiles are usually taken as a signature of accretion funnel flows passing on the line of sight \citep{Edwards94}.
        Fig.~\ref{fig:columnsprof} shows the temporal sequence of residual line profiles. IPC profiles are seen in the H$\gamma$ line  between phase 0.26 and 0.34, which is shifted by about 0.2 in phase with the photometric minimum.
        We provide a larger view of those profile on Fig.~\ref{fig:Hgphase}.
        We also notice that the last three spectra of the series exhibit Balmer line profiles with a strongly depressed blue wing, which suggests enhanced outflow activity. This appears to be a transient event occurring over a few nights. 
        
        We performed a periodogram analysis in each 2 \kms-wide velocity channel across the line profiles,  which we refer to as 2D periodograms. These are shown in Fig.~\ref{fig:reslines}. A periodic modulation is seen in the central absorption component of the \ha\ profile, and in the red wing of the \hb\ and \hg\ line profiles. The maximum periodogram power occurs at a frequency around 0.23 d$^{-1}$ (4.3 d), which is consistent with the stellar rotation period.
        The false alarm probability (FAP) of the peak in the central absorption of the  H$\alpha$ line and in the redshifted wings of the \hb\ and \hg\ profiles is about 10$^{-2}$. This indicates that various parts of the Balmer line profiles are intensity modulated on the stellar rotation period.

        \begin{figure}
            \centering
            \begin{subfigure}{.24\textwidth}
                \includegraphics[width=\textwidth]{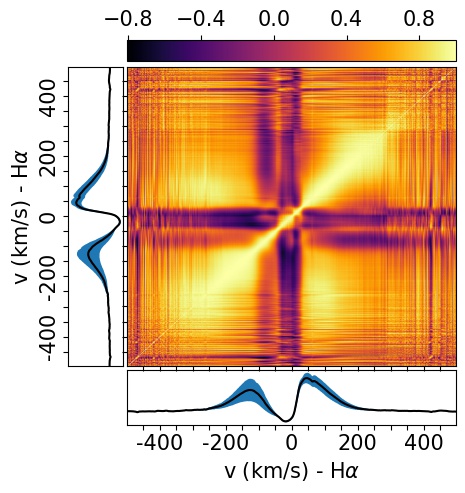}
                \caption{H$\alpha$}
                \label{fig:acmHa}
            \end{subfigure}
            \begin{subfigure}{.24\textwidth}
                \includegraphics[width=\textwidth]{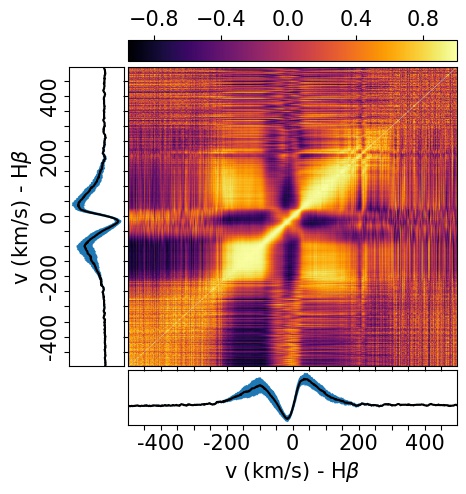}
                \caption{H$\beta$}
                \label{fig:acmHb}
            \end{subfigure}
            \begin{subfigure}{.24\textwidth}
                \includegraphics[width=\textwidth]{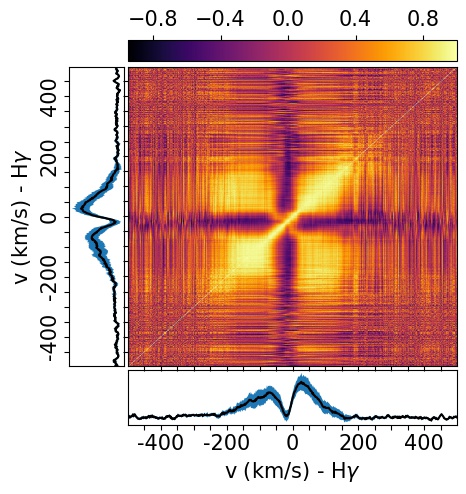}
                \caption{H$\gamma$}
                \label{fig:acmHg}
            \end{subfigure}
            \begin{subfigure}{.24\textwidth}
                \includegraphics[width=\textwidth]{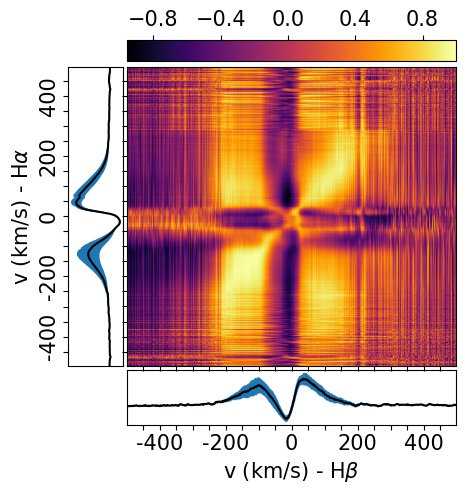}
                \caption{H$\alpha$ vs. H$\beta$}
                \label{fig:acmHaHb}
            \end{subfigure}
            \begin{subfigure}{.24\textwidth}
                \includegraphics[width=\textwidth]{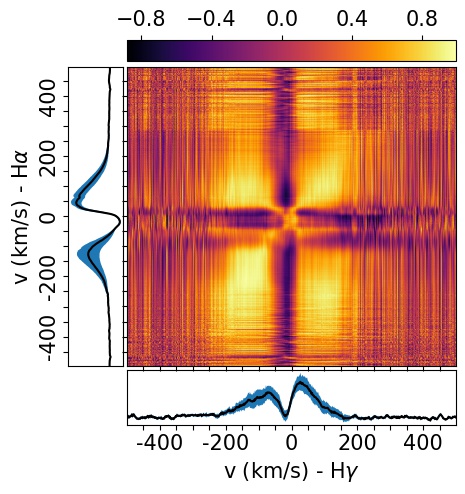}
                \caption{H$\alpha$ vs. H$\gamma$}
                \label{fig:acmHaHg}
            \end{subfigure}
            \begin{subfigure}{.24\textwidth}
                \includegraphics[width=\textwidth]{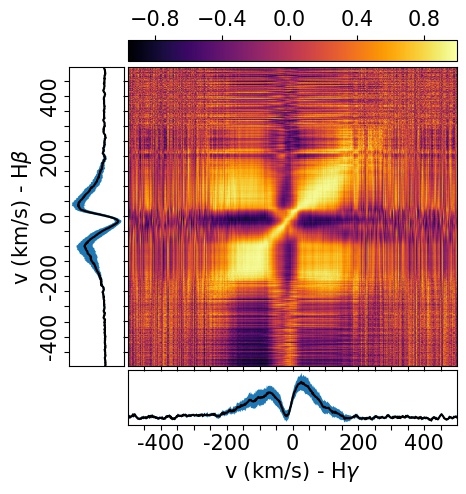}
                \caption{H$\beta$ vs. H$\gamma$}
                \label{fig:acmHbHg}
            \end{subfigure}
            \caption{Correlation matrix of residual Balmer lines. Light yellow represents a strong correlation, dark purple a strong anti-correlation, and orange means no correlation. The bottom and left inserts in each panel show the mean residual line profile (black) and its variance (blue) for each axis.}
            \label{fig:acmBlines}  
        \end{figure}
        
        The correlation matrix is a powerful tool to investigate the relationship between intensity variations occurring over different velocity channels within the line profiles. 
        It consists of computing a linear correlation coefficient between each velocity channel of a pair of lines.
        A strong correlation reveals a link between the two velocity channels, which points to a common physical origin for their variability.   
        Figure \ref{fig:acmBlines} shows the autocorrelation matrices, meaning  the correlation matrix between a line and itself, for the H$\alpha$, H$\beta$, and H$\gamma$ line profiles.
        The three lines show the same behavior, namely a strongly correlated variability within the red wing, and a similar behavior over the blue wing. 
        The central absorption seems slightly anticorrelated with the rest of the profile.
        Fig.~\ref{fig:reslines} shows that the central absorption component of the H$\alpha$ profile hardly varies, while those seen in the H$\beta$ and H$\gamma$ profiles exhibit more significant variability.  
        The cross-correlation matrices between the Balmer lines, also shown in Fig.~\ref{fig:acmBlines}, display the same pattern as the  autocorrelation matrices. This suggests the three Balmer lines form under similar conditions in the immediate environment of the star.  
        
\begin{figure}
            \centering
            \includegraphics[width=.45\textwidth]{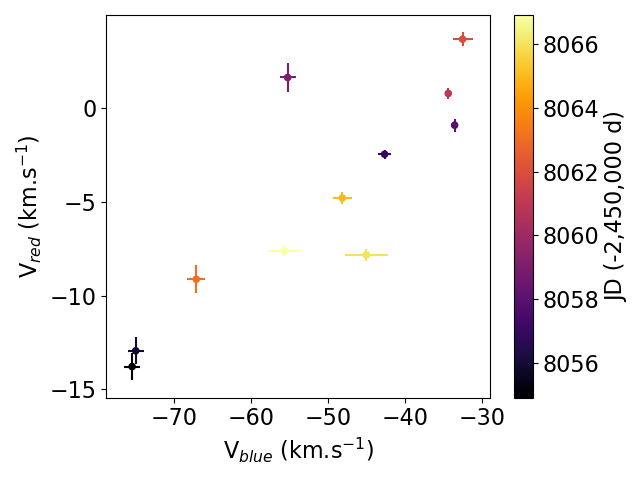}
            \caption{Radial velocity of the two absorption components seen in the \ha\ line profile. The color scale indicates the Julian Date of observation.}
            \label{fig:VredVblue}
        \end{figure}

        AA Tau \citep{Bouvier03} and LkCa 15 \citep{Alencar18} exhibit similar Balmer line profiles as V807 Tau. This prompted us to attempt a Gaussian decomposition of the H$\alpha$ profile for V807 Tau, as was previously done for these other young stellar objects. We assume the profile is a superposition of  
        a broad central emission and two absorption components. The profile decomposition is illustrated in Fig.  \ref{fig:HaDecomp} and the radial velocities of each component are listed in Table~\ref{tab:rad_vel}.
        The global fit yields a central emission that is slightly blueshifted, with V$_{em}$ $\sim$ -20 \kms, a blueward asymmetry sometimes ascribed to disk occultation  \citep{Muzerolle01, Alencar00}. 
        The velocity of the bluer absorption component varies between -75 and -32 \kms, while the velocity of the redder component, still lying mostly on the blue side of the line profile, ranges from -14 to +4 \kms. 
        As shown in Fig.~\ref{fig:VredVblue}, the radial velocities of the two absorption components seem to correlate. The linear dependence is quite similar to that previously reported for the blue and red  absorption components seen in the \ha\ line profiles of AA Tau and LkCa 15, a result that was interpreted as evidence for magnetospheric inflation \citep[][]{Bouvier03}. 
        This phenomenon occurs when the truncation radius is not located at the corotation radius.
         Differential rotation along the  accretion funnel flow forces the magnetic field lines to inflate, until they open and eventually reconnect to restore the initial configuration  \citep{Goodson97}.
        The reconnection process induces magnetospheric ejections \citep{Zanni13}, which might be responsible for the blueshifted absorptions seen during the last three nights of observation.

        \begin{table*}	
            \small
                  \centering
                  \begin{tabular}{l|ll|ll|lll|llll}
                      \hline
                      &\multicolumn{2}{| c }{Photosphere}&\multicolumn{2}{| c }{\He~I}&\multicolumn{3}{| c }{H$\alpha$} & \multicolumn{4}{| c }{EW}\\
                      HJD & \vrad & $\sigma$ \vrad & \vrad & $\sigma$ \vrad & V$_{em}$ & V$_{blue}$ & V$_{red}$ & H$\alpha$ & H$\beta$ & H$\gamma$ & He~I (NC)\\
                      (2450000+) & \multicolumn{2}{| c }{(\kms)}  & \multicolumn{2}{| c }{(\kms)}& \multicolumn{3}{| c }{(\kms)}  &  \multicolumn{4}{| c }{(\AA)}\\
                      \hline
                      8054.90048 & 17.1 & 0.3 & 8.8 & 0.4 & -28.3 & -75.4 & -13.8 & 10.7 & 4.3 & 3.1 & 0.14 \\
                      8055.89091 & 17.2 & 0.3 & 4.3 & 0.4 & -16.0 & -75.0 & -12.9 & 11.2 & 4.4 & 3.5 & 0.17 \\
                      8057.00939 & 17.2 & 0.2 & 5.7 & 0.4 & -18.3 & -42.6 & -2.5 & 9.1 & 3.1 & 2.2 & 0.17 \\
                      8057.99515 & 16.8 & 0.3 & 8.4 & 0.4 & -16.9 & -33.6 & -0.9 & 12.2 & 5.0 & 3.6 & 0.21 \\
                      8059.05448 & 17.0 & 0.3 & 8.8 & 0.3 & -31.6 & -55.2 & 1.6 & 13.2 & 5.0 & 3.3 & 0.19 \\
                      8060.95421 & 17.4 & 0.2 & 2.5 & 0.4 & -27.3 & -34.4 & 0.8 & 11.7 & 3.7 & 2.5 & 0.18 \\
                      8061.95085 & 16.9 & 0.3 & 5.8 & 0.3 & -20.9 & -32.5 & 3.6 & 9.9 & 3.6 & 3.0 & 0.24 \\
                      8063.08059 & 17.8 & 0.3 & 8.9 & 0.3 & -18.7 & -67.1 & -9.1 & 14.2 & 5.2 & 3.8 & 0.2 \\
                      8064.99530 & 17.3 & 0.2 & 3.0 & 0.5 & -10.3 & -48.1 & -4.8 & 5.2 & 1.7 & 1.5 & 0.15 \\
                      8065.99958 & 17.1 & 0.3 & 5.5 & 0.4 & 6.7 & -45.0 & -7.8  & 4.5 & 2.0 & 1.4 & 0.17 \\
                      8066.91484 & 16.2 & 0.3 & 9.0 & 0.3 & -25.1 & -55.7 & -7.6 & 6.2 & 2.9 & 3.0 & 0.21 \\
                      \hline
                  \end{tabular}
                  \caption{Radial velocities listed for photospheric lines, the He~I narrow emission component, and for each of the Gaussian components of the \ha\ profile decomposition (see text). The EWs are also listed for Balmer lines and the He~I narrow component.}
                  \label{tab:rad_vel}
        \end{table*}


    \subsubsection{He~I 587.6 nm}
    \label{subsec:heI}

        The He~I 587.6 nm line profiles seen in the ESPaDOnS spectra are displayed in Figs. \ref{fig:reslines} and \ref{fig:columnsprof}.  
         The line profile exhibits broad (FWHM$\sim$130 \kms) and narrow (FWHM$\sim$30 \kms) components.
        The narrow component (NC) is thought to form at least partly in the accretion post-shock at the stellar surface \citep{Beristain01}. It displays strong variability and the 2D periodogram shown in Fig.~\ref{fig:reslines} suggests it is modulated at the 
         stellar rotation period, with a FAP$\sim 10^{-3}$.
        Fig.~\ref{fig:columnsprof} illustrates the periodic intensity variations of the He~I line profile. 
        The peak intensity of the narrow component of the He~I line profile increases continuously from phase 0.03 to phase 0.34, and decreases again thereafter. 

\begin{figure}
            \centering
            \includegraphics[width=.49\textwidth]{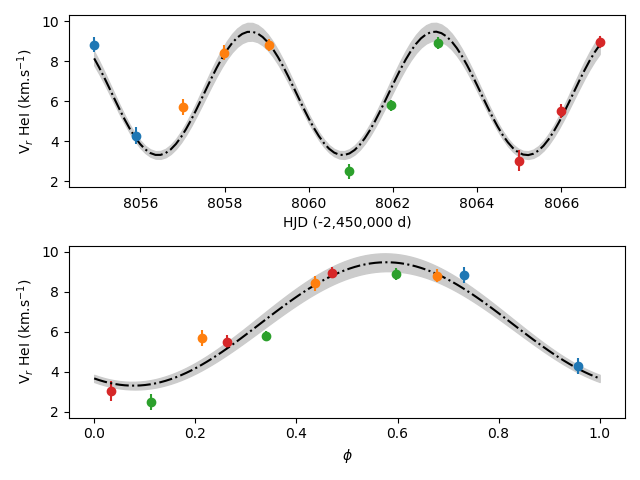}
            \caption{Radial velocity curve of the He~I line narrow component plotted as a function of Julian date ({\it top}) and rotational phase ({\it bottom}). The fit by the model fully described in Sec~\ref{subsec:heI} is shown ({\it dash-dotted curve}) together with its 1$\sigma$ uncertainty ({\it gray area}). The color scale indicates the rotational cycle.}
            \label{fig:vrhe}
        \end{figure}

        We performed a two-Gaussian decomposition of the line profile in order to isolate the narrow and broad components. 
        We thus derived the centroid velocity of the narrow component, which is shown in Fig.~\ref{fig:vrhe}.
        The NC is redshifted by $\sim$+6 \kms\ on average over the whole data set. 
        The radial velocity displays variations with an amplitude of about 3 \kms\ and is modulated at the stellar period. The maximum velocity occurs from phase 0.5 to 0.7. 
        The radial velocity curve of the He~I NC results from a combination of two effects:  the intrinsic flow velocity in the post-shock region, V$_{flow}$, and the radial velocity shift of the hot spot due to stellar rotation, $\delta$V$_r$. 
        
        We develop a geometrical model to account for the periodic radial velocity variations of He~I NC. 
        We first consider the intrinsic velocity of the gas in the post shock region, going toward the star, V$_{flow}$.
        We used spherical coordinates to describe the velocity vector, with $\varphi$ the rotational phase ($\in$ [0 ; 2$\pi$]), $\theta$ the colatitude of the spot, and V$_{flow}$ the radial velocity of the infalling material.
        The cartesian coordinates are thus described by:
        \begin{equation}
            \begin{cases}
                x = -V_{flow} \sin \theta \cos \varphi \\
                y = -V_{flow} \sin \theta \sin \varphi \\
                z = -V_{flow} \cos \theta ,
            \end{cases}
        \end{equation}
        where we assumed we observe the system along the x-axis in the case of an edge-on system (i=90\deg). 
        Taking into account the actual stellar inclination $i$, 
        we applied a rotation around the y axis by an angle $\alpha$ = i - 90\deg,
        yielding:
        \begin{equation}
            \begin{cases}
                x' = x \cos \alpha + z \sin \alpha \\
                y' = y \\
                z' = -x \sin \alpha + z \cos \alpha ,
            \end{cases}
        \end{equation}
        where $x'$ is thus V$_{flow}$ projected on the line of sight.
        We then consider the velocity shift due to stellar rotation, given by: 
        \begin{equation}
            \delta V_r = -v\sin i \sin \theta \sin \varphi ,
        \end{equation} 
        where we assume the spot faces the observer at phase 0.5 (\textit{i.e.}, $\varphi$ = $\pi$).
        If the spot is located at another phase shifted by $\delta \varphi$, we simply apply the transformation $\varphi '$ = $\varphi$ - $\delta \varphi$.
        The apparent radial velocity of the He~I NC is thus given by V$_r$(He~I NC) = x' + $\delta$V$_r$.
        
        Using the \vsini\ and inclination derived in Sect.~\ref{subsec:stellarparam}, and a hot spot latitude of 70\deg\ derived in Sect.~\ref{subsec:resspectropola}, we derived  V$_{flow}$ = 9.5 $\pm$ 0.7 \kms, which is consistent with accretion shock simulations by \cite{Matsakos13}. The model fit is shown in Fig.~\ref{fig:vrhe}. We further derive that the hot spot faces the observer at \phirot = 0.55 $\pm$ 0.07.
        This is consistent with the photometric minimum, and thus indicates that the accretion shock and the occulting structure are located at the same azimuth, as expected for a corotating funnel flow, and do not vary over more than 5 months.

        Equivalent width (EW) measurements for the He~I NC are summarized in Table~\ref{tab:rad_vel}.
        The largest EW occurs at phase 0.34, which correspond to the phase where the IPC profile is seen in the  H$\gamma$ line profile. 


    \subsubsection{Other lines}

        Other lines in the spectra of V807 Tau may provide clues to the star disk interaction region. We review them here briefly. 
    
\begin{figure*}
            \centering
            \begin{subfigure}{.3\textwidth}
                \includegraphics[width=\textwidth]{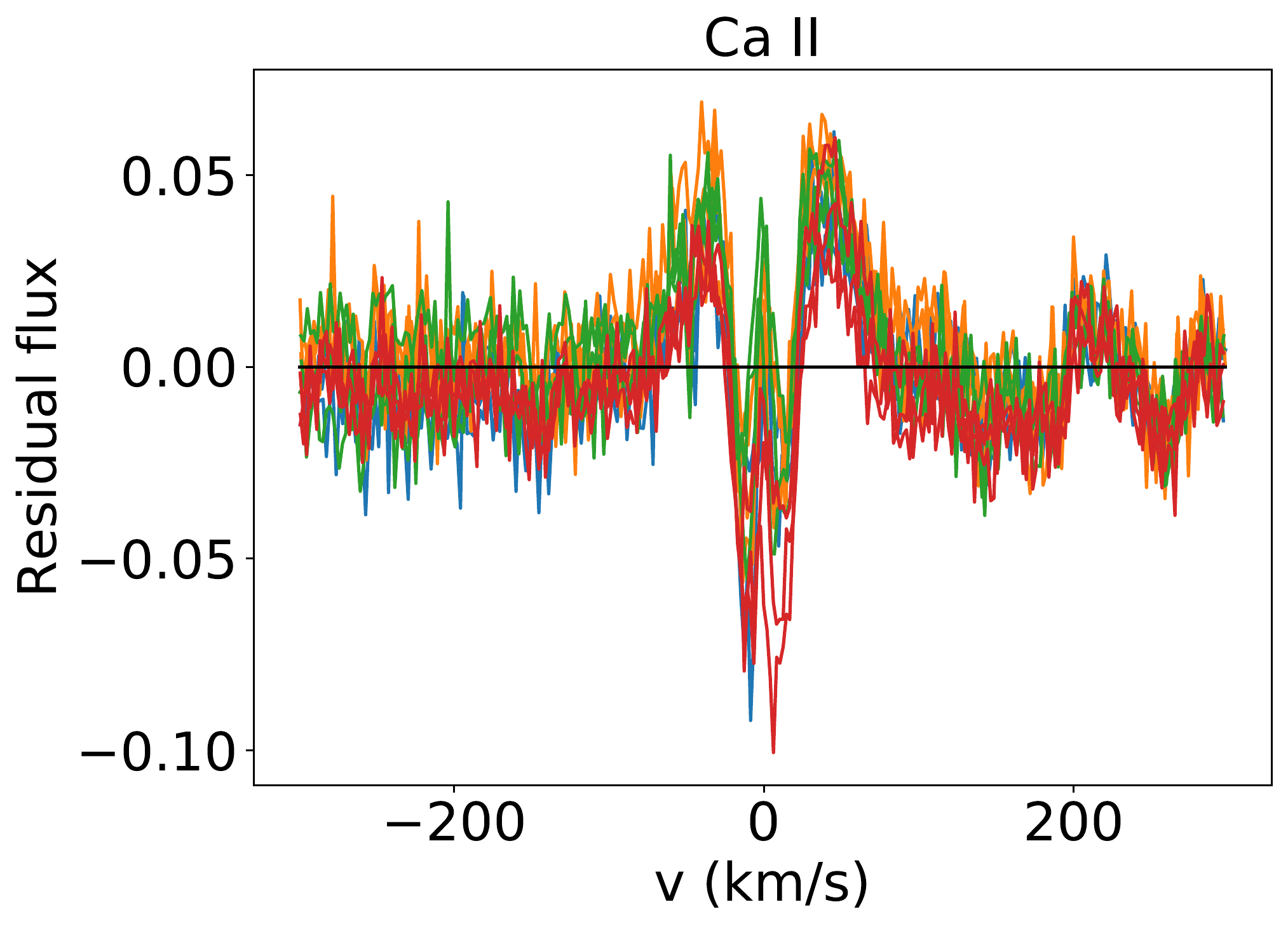}
                \label{fig:CaSup}
            \end{subfigure}
            \begin{subfigure}{.3\textwidth}
                \includegraphics[width=\textwidth]{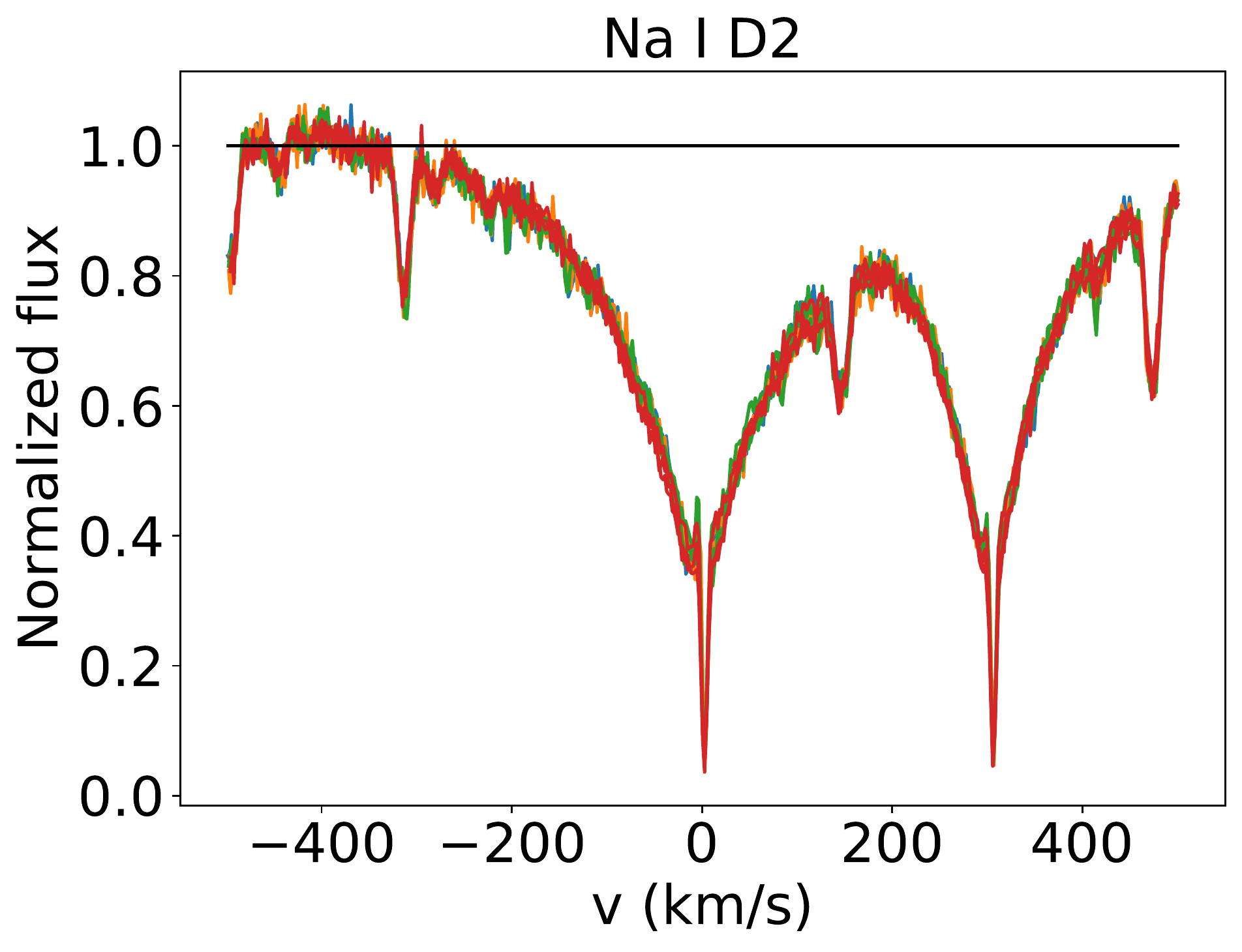}
                \label{fig:NaSup}
            \end{subfigure}
            \begin{subfigure}{.3\textwidth}
                \includegraphics[width=\textwidth]{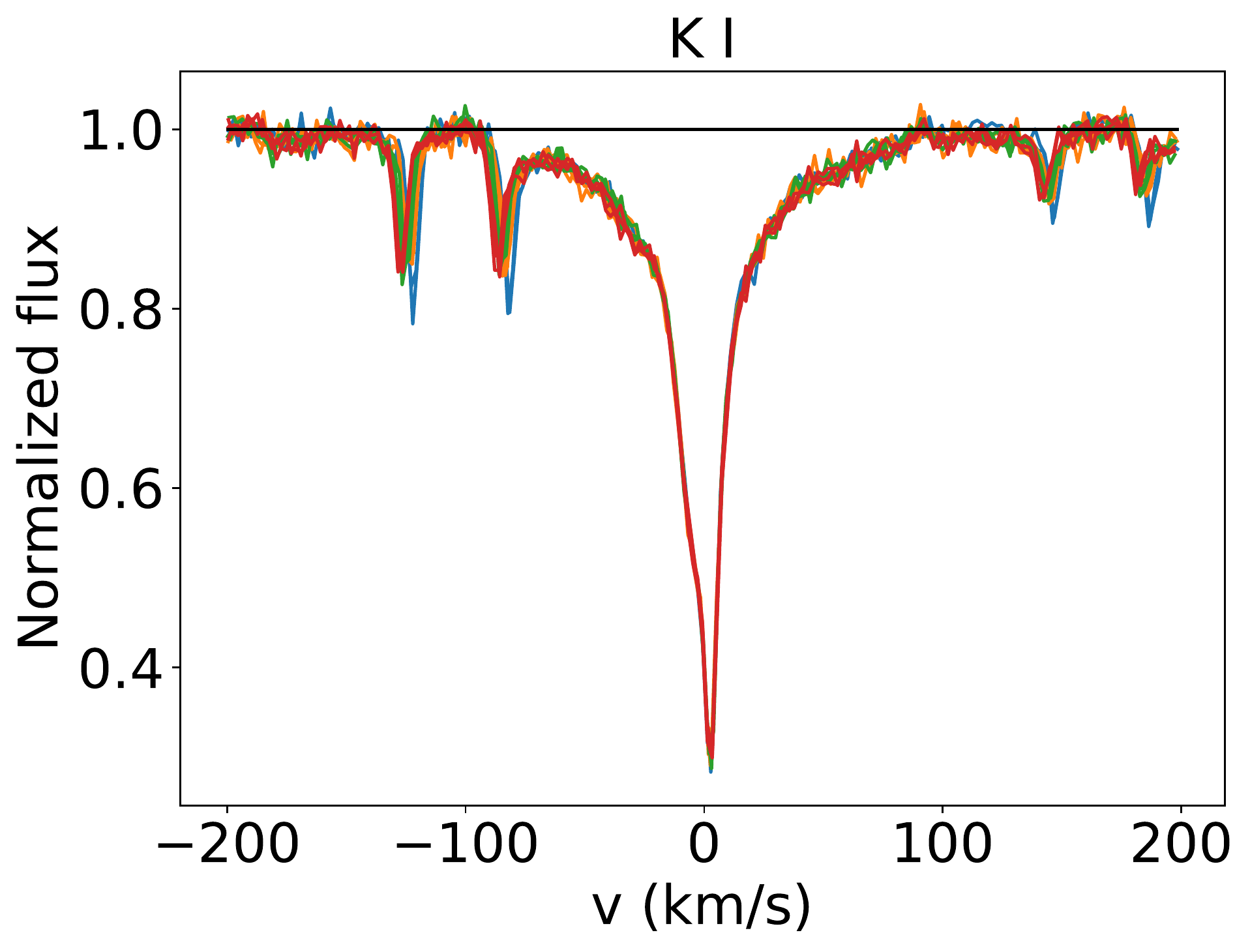}
                \label{fig:KSup}
            \end{subfigure}
            \caption{{\it From left to right:} The 854 nm residual line profile of Ca~II IRT, and the NaI D2 and KI (770 nm) line profiles. The color code indicates the rotation cycles.}
            \label{fig:otherlines}
        \end{figure*}

        The 854 nm residual line of Ca~II infrared triplet (IRT) is shown in Fig.~\ref{fig:otherlines}. It exhibits the same double peaked shape as the Balmer lines except for a central reversal in emission that might be of chromospheric origin.
        This line is weaker and thus more difficult to analyze. Nevertheless, it exhibits significant variability, especially the central part of the profile.
        The periodogram analysis does not reveal any significant periodicity.

        The NaI D lines are seen deeply in absorption in the spectra of V807 Tau. 
        They are shown in Fig.~\ref{fig:otherlines} where we set the stellar rest velocity on the D2 line at 588.9 nm. No significant variability is seen in these profiles. In particular, no sign of episodic outflows, as reported in these lines for another young system HQ Tau \citep[see][]{Pouilly20}, is observed here.
        The deep, narrow, and stable absorption component seen at the stellar rest velocity is consistent with  interstellar absorption from the Taurus cloud (V$_{helio}$ $\sim$ 18 \kms, \cite{Pascucci15}).



        The KI line at 770 nm often shows the same structure as Na I lines \citep{Pascucci15}.
        The profiles are shown in Fig.~\ref{fig:otherlines}.
        They are quite stable, exhibiting deep photospheric absorption, and an interstellar absorption component at the center of the line. 
        


        
    \subsubsection{Mass accretion rate}
    \label{subsec:macc}
        We computed V807 Tau's mass accretion rate (\macc) using the Balmer residual lines' EW summarized in Table~\ref{tab:rad_vel}.
        We thus derived the line flux using $F_{line} = F_{0}\cdot EW \cdot 10^{-0.4m_{\lambda}}$ with $F_{0}$ the reference flux, $EW$ the line EW, and $m_{\lambda}$ the extinction-corrected magnitude in the selected filter.
        We used the photometry from CrAO described in Sect.~\ref{sec:obs}, converting the colors from the Johnson to the Cousins system using the relationships given by \cite{Fernie83}.
        The line luminosity is computed as $L_{line} = 4 \pi d^{2} F_{line}$, with $d$ the distance to the object, and we used the relationships between $L_{line}$ and the accretion luminosity 
        provided by \cite{Alcala17}.
        The mass accretion \macc\ is obtained from:
        \begin{equation}
            L_{\rm acc} = \frac{{\rm G} {\rm M}_{\star}\dot{M}_{\rm acc}}{{\rm R}_{\star}}\bigg[1-\frac{{\rm R}_{\star}}{{\rm R}_{t}}\bigg],
            \label{eq:macc}
        \end{equation}
        using $R_{t} = 5 R_{\star}$, a typical truncation radius for CTTSs \citep{Bouvier07a}.
        We derived \macc(H$\alpha$) = 1.0 $\pm$ 0.2 10$^{-9}$ \msunyr, \macc(H$\beta$) = 5$\pm$ 2 10$^{-10}$ \msunyr, \macc(H$\gamma$) = 7 $\pm$ 3 10$^{-10}$ \msunyr.
        We thus adopt <\macc> = 7 $\pm$ 2 10$^{-10}$ \msunyr as the mean mass accretion rate onto the star during our observations.
        
\begin{figure}
            \centering
            \includegraphics[width=.45\textwidth]{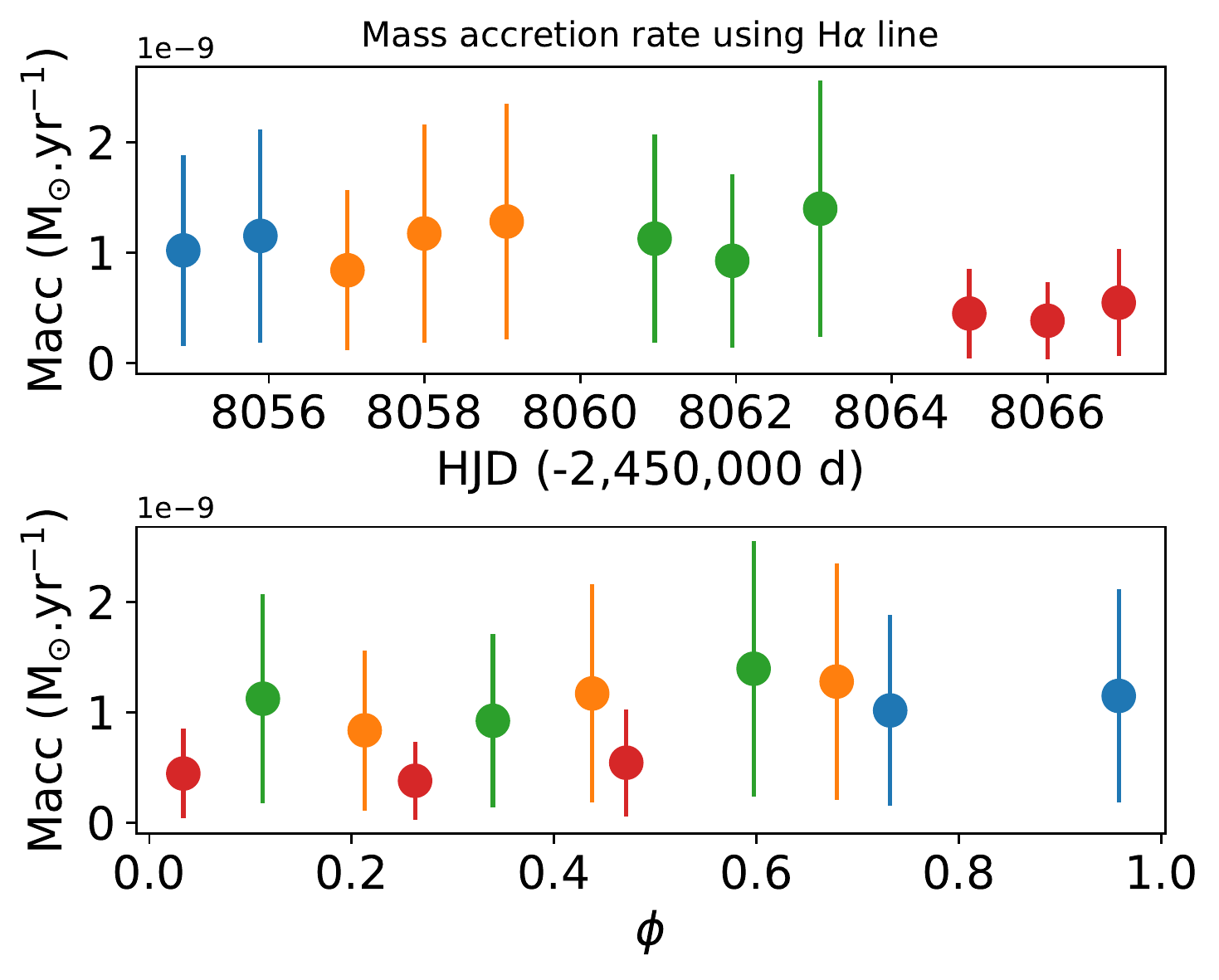}
            \caption{Mass accretion rate computed from the H$\alpha$ line is plotted as a function of Julian date ({\it top}) and rotation phase ({\it bottom}).}
            \label{fig:macc}
        \end{figure}

        As shown in Fig.~\ref{fig:macc}, the \macc\ variations during our observing run are small. They follow the EW of the Balmer lines (see Table~\ref{tab:rad_vel}). \macc\  from H$\alpha$ line fluctuates by about 20\%  around a value of 10$^{-9}$ \msunyr, being stable within measurement uncertainties, except for the last three observations where it decreased by about a factor of two. Removing the last three measurements, a periodogram analysis yields weak evidence for the modulation of the accretion rate along the rotational cycle, as expected from a corotating accretion funnel flow, with a maximum around phase 0.6.  

    \subsection{Spectropolarimetric analysis}
    \label{subsec:resspectropola}
\begin{figure*}[t]
            \centering
            \includegraphics[width=.99\textwidth]{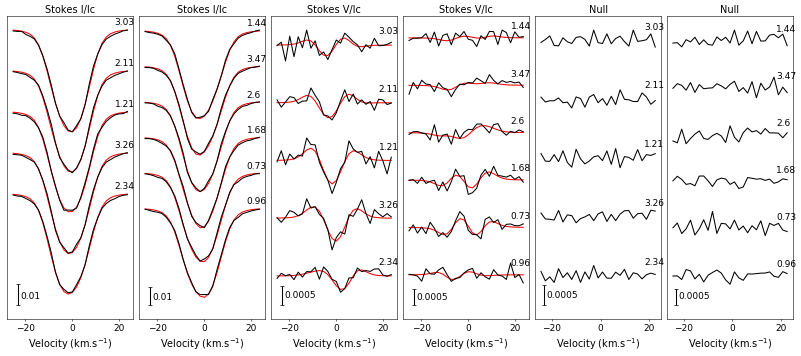}
            \caption{V807 Tau's Stokes I {\it (left)}, V {\it (middle)}, and Null {\it (right)} profiles (black) ordered by increasing fractional phase,  and the fit produced by the ZDI analysis (red). The decimal part of each number next to each profile indicates the rotational phase. }
            \label{fig:stokes_i_v}
        \end{figure*}

        We used the ESPaDOnS data set in polarimetric mode to study the magnetic field properties of V807 Tau.
        We first computed the Stokes I (unpolarized) and V (circularly polarized) photospheric profiles using the Least Square Deconvolution method (LSD - \cite{Donati97}).
        This method increases the S/N of Zeeman signatures by extracting them from as many photospheric lines as possible (about 6000 lines in this study).
        We used the mean wavelength, intrinsic line depth, and Landé factor of 640 nm, 0.2. and 1.2, respectively \citep{Pouilly20}.
        The photospheric lines were selected from the same VALD line list as in Sect. \ref{subsec:stellarparam}, over the 490-990 nm wavelength range.
        We removed emission, telluric, and heavily blended lines to produce a suitable line mask and compute LSD Stokes profiles.
        The profiles are shown in Fig.~\ref{fig:stokes_i_v}. A clear magnetic signal is detected in the Stokes V profiles, indicative of a significant large-scale magnetic field.

\begin{figure}[t]
            \centering
            \includegraphics[width=.45\textwidth]{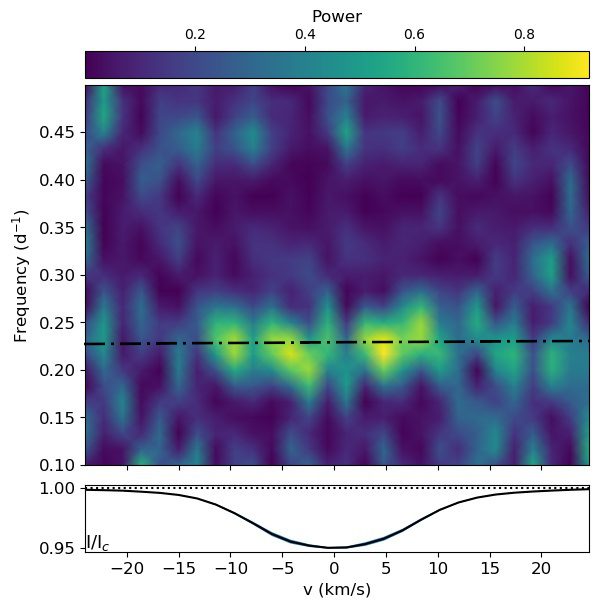}
            \caption{2D periodogram of LSD Stokes I photospheric profiles. Light yellow represents the highest power of the periodogram. The mean profile (black) and its standard deviation (blue) are represented in the bottom panel. The dashed-dotted line indicates the rotational period of the star}
            \label{fig:stokes_i_periodo}
        \end{figure}

        We performed a 2D periodogram analysis on the LSD Stokes I profiles. The periodogram is 
        shown in Fig.~\ref{fig:stokes_i_periodo} and reveals a clear periodicity (FAP $\sim$ 10$^{-3}$) consistent with the rotation period  over the whole profile.
        The same analysis was done on the Stokes V profiles and did not reveal any periodicity, even though it seems to smoothly vary in phase. 

\begin{figure}[t]
            \centering
            \includegraphics[width=.45\textwidth]{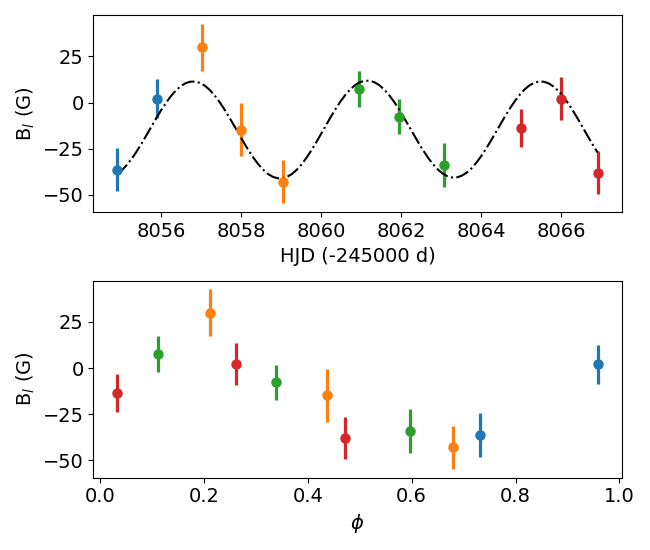}
            \caption{Surface-averaged longitudinal magnetic field is plotted as a function of date in the top panel and as a function of rotational phase in the bottom panel, using the ephemerides described in Sect. \ref{subsec:resphoto}. The dash-dotted curve in the top panel is the best sinusoid fit. The colors indicate successive rotation cycles. }
            \label{fig:bl25}
        \end{figure}

\begin{figure*}[t]
            \centering
            \includegraphics[width=.95\textwidth]{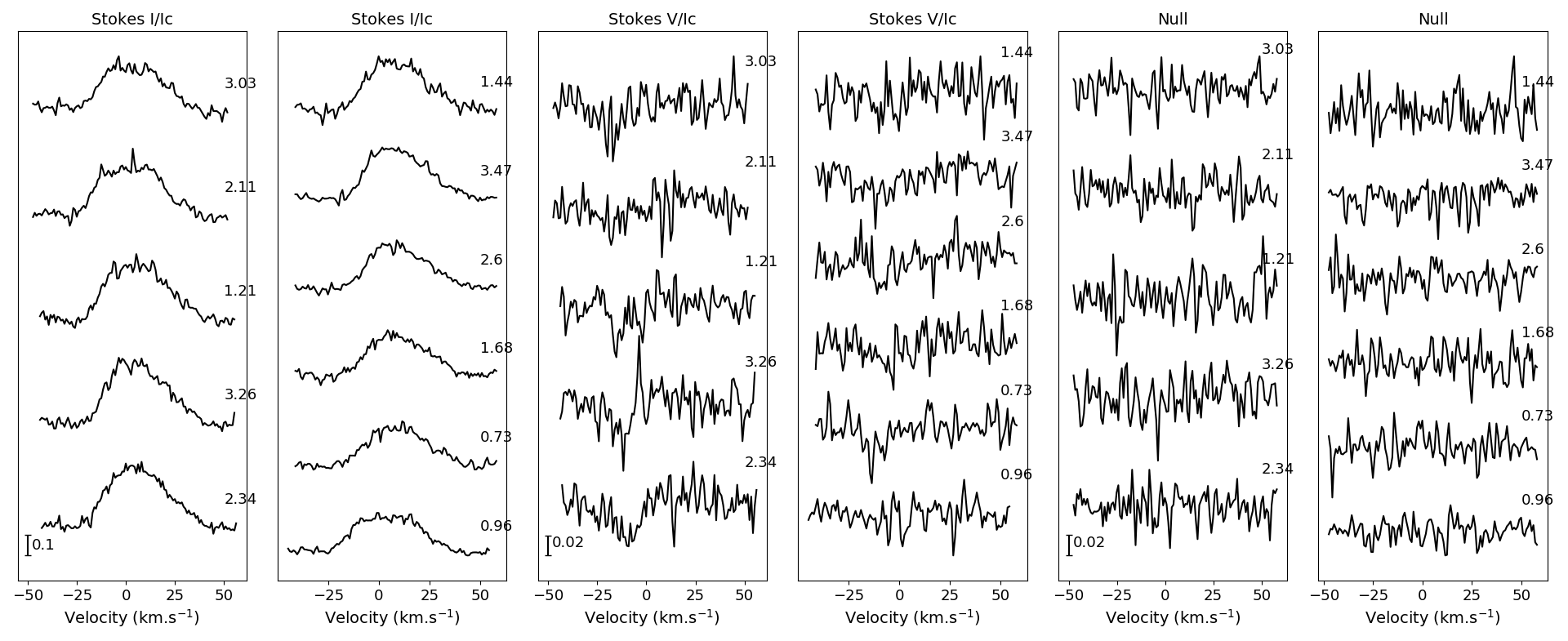}
            \caption{He I 587.6 nm narrow component \textit{(left)}, circularly polarized \textit{(middle)}, and null signal \textit{(right)}, ordered by increasing phase.}
            \label{fig:hezdifit}
        \end{figure*}

\begin{figure}[t]
            \centering
            \includegraphics[width=.45\textwidth]{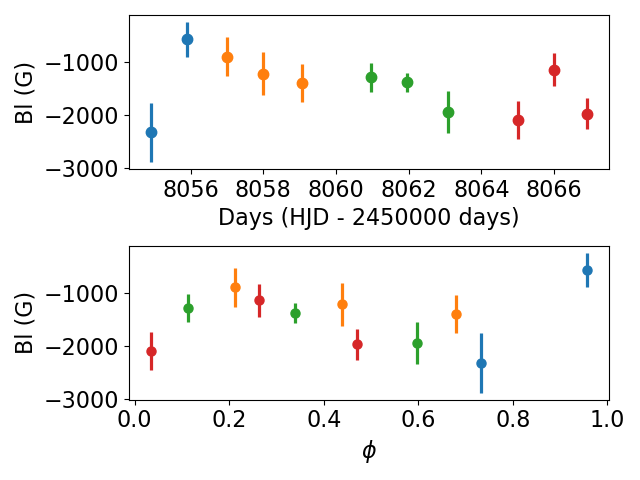}
            \caption{Surface-averaged longitudinal magnetic field computed from the He I line as a function of the date \textit{(top panel)} and rotational phase\textit{(bottom panel)}. The colors indicate successive rotation cycles.}
            \label{fig:blhe}
        \end{figure}

        The Stokes I and V LSD profiles yield an estimate of the average longitudinal surface magnetic field component ($B_l$), using the relationship \citep{Donati97, Wade01}:

        \begin{equation}
            B_l = -2.14 \times 10^{11} \times \frac{\int vV(v)dv}{\lambda g c \int (1-I(v))dv},
            \label{eq:bl}
        \end{equation}
        where $B_l$ is expressed in Gauss, $v$ is the velocity relative to line center, $\lambda$ is the mean wavelength in nm, and $g$ the Land\'e factor chosen for the LSD computation.
        The analysis was done over the range $\pm$ 25 \kms\ around the stellar radial velocity, \vrad $\sim$17 \kms, in order to avoid  integration over the continuum which would increase the noise without adding signal.
        We propagated the uncertainties over a trapezoidal integration to produce the error bars. The resulting $B_l$ curve is shown in Fig.~\ref{fig:bl25}.
        The $B_l$ component is clearly modulated and a sinusoidal fit reveals a period of 4.3 $\pm$ 0.2 d, consistent with the stellar period. We derive a mean <$B_l$> = -13 $\pm$ 3 G, with extrema reaching about +30 G and -40 G, at phases $\sim$0.2 and $\sim$0.6, respectively. 
        Following the same method, we also derived B$_{l}$ within the He~I NC profiles shown in Fig.~\ref{fig:hezdifit}, thought to form in the post-shock region.
        The results are shown in Fig.~\ref{fig:blhe}. The B$_l$ component of the HeI line is modulated in the same way as the B$_l$ component of the photospheric LSD profiles, but is much stronger, with an intensity reaching an absolute minimum of about -2kG around phase 0.7, consistent with the accretion shock location. 

        Using the photospheric LSD profiles, we performed a full Zeeman-Doppler imaging (ZDI) analysis \citep{Donati11, Donati12} with the Python algorithm described in \cite{Folsom18}.
        We derived the magnetic field topology in two steps. 
        We first built a Doppler image, which consists in iteratively adding dark and bright features on a uniformly bright stellar disk till the Stokes I profiles are reproduced.
        The algorithm assumes a local Voigt profile for each cell at the stellar surface and minimizes the $\chi^2$ while maximizing the entropy to adjust the brightness.
        We defined the Voigt parameters, meaning the Gaussian and Lorentzian widths, by fitting a Voigt profile to a slow rotator of the same spectral type. 
        We chose the K6 type star HD88230, with \vsini = 1.8 \kms, yielding a Gaussian and a Lorentzian width of 2.66 \kms \ and 0.604 $\times$ 2.66 \kms, respectively.

\begin{figure}[t]
            \centering
            \includegraphics[width=.45\textwidth]{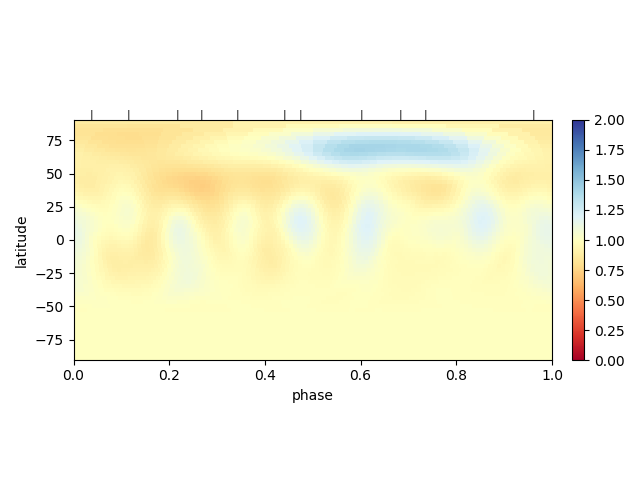}
            \caption{Brightness map reconstructed from the LSD Stoke I profiles. Blue represents a brighter region and red a darker region. The mean brightness of the stellar surface is yellow. The ticks above the map are the phase of observation.}
            \label{fig:bm}
        \end{figure}

        We first performed the DI analysis using the stellar parameters we derived in previous sections in order to obtain a first target $\chi^2$.
        Then we adjusted the stellar parameters by computing DI maps on several grids of parameters around the estimated values and identify the minimal $\chi^2$ and maximal entropy of each grid.
        The grids consist of 2D maps in the parameter space for all the free parameters such as the period, differential rotation,  \vsini, \vrad, and inclination angle.
        The period versus differential rotation grid (P vs. d$\Omega$) does not allow us to constrain those parameters; we thus chose to fix the period at the value found in Sect.~\ref{subsec:resphoto}, P = 4.38 d, and set d$\Omega$ = 0.
        The other grids yielded \vsini = 11.2 $\pm$ 0.3 \kms, \vrad = 15.8 $\pm$ 0.2 \kms\ and i = 53 $\pm$ 5\deg. 
        These values are  within 3 $\sigma$ of the estimates we derived above for these parameters.
        The resulting fits of the Stokes I profiles are shown in Fig.~\ref{fig:stokes_i_v} and the corresponding brightness map is displayed in Fig.~\ref{fig:bm}.
        The Doppler map reveals a bright feature located at a latitude of $\sim$ 70\deg \ extending from phase 0.4 to 0.8, consistent with the longitude at which the $B_l$ curve's maximum absolute value occurs. It also corresponds to the phase at which the He I line intensity is maximum and to the longitude of the hot spot we derived above by modeling the radial velocity variations of the HeI NC line profile. 
        A darker region also appears at mid-latitudes, which extends from phase 0.1 to 0.4, consistent with the cool spot longitude inferred above from the photospheric \vrad\ curve (Fig.~\ref{fig:rvcurve}).
        
\begin{figure}[t]
            \centering
            \includegraphics[width=.45\textwidth]{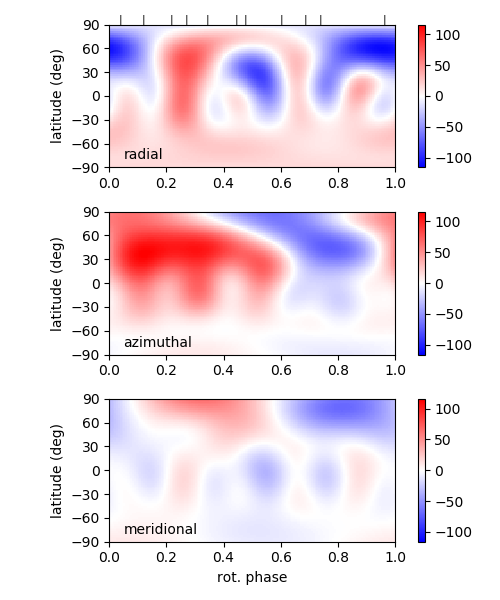}
            \caption{Radial (\textit{top}), azimuthal (\textit{middle}), and meridional (\textit{bottom}) magnetic field maps. Redder region are positive field strength and bluer are negative (in G). The ticks above the top map are the phases of observation.}
            \label{fig:mm}
        \end{figure}

        We then computed the Zeeman Doppler image, which consists in fitting the Stokes V profiles  starting from the computed brightness map in order to derive the magnetic field topology by adjusting the spherical harmonic components detailed in \cite{Donati06}.
        Fitting the circularly polarized profile beyond the l=5 spherical harmonic order does not change the magnetic maps; we thus limited the spherical harmonics to this $l$ value.
        The resulting surface maps for the radial, azimuthal, and meridional magnetic components are shown in Fig.~\ref{fig:mm}.

        The ZDI analysis suggests a large-scale field that is mostly poloidal (69~\% of the total magnetic energy). 
        A significant dipolar component (47~\% of the poloidal energy)  reaches 47~G at the pole, located at phase 0.7 and a latitude of about 50\deg. 
        The quadrupolar and octopolar components represent 21~\% and 13~\% of the poloidal energy, respectively.
        Locally, the magnetic field reaches a strength up to 135 G, and its unsigned field strength averaged over the surface amounts to B$_{mean}$ = 50~G.
        The magnetic topology of V807 Tau presents a total axisymmetry of about 23~\% ({\it i.e.,} the percentage of the total magnetic energy in components that are symmetric about the rotation axis, meaning for the spherical harmonic degree m = 0).
        The axisymmetry of the poloidal components ({\it i.e.,} considering only the $\alpha_{l, m=0}$ and $\beta_{l, m=0}$ components of the spherical harmonic expansion) represents about 8~\%.

\section{Discussion}
\label{sec:discussion}
    Following the identification of V807 Tau as a stable periodic dipper from its K2 light curve, we selected 
    this system to be monitored by CFHT/ESPaDOnS in spectropolarimetric mode.  
    
    V807 Tau is a 0.72 \msun\ CTTS of spectral type K7, and is fully convective according to PMS evolutionary models.  
    The dipper-like K2 light curve exhibits a clear period of 4.38 d, with a dip amplitude amounting to about 10\%.
    If those dips are due to obscuration by circumstellar dust located close to the corotation radius, the period ought to reflect the stellar rotational period. This is confirmed by the spectroscopically derived radial velocity variations induced by surface spots that we find to be periodic at the same period.
    This is also the case for the modulation of the longitudinal component of the surface magnetic field we measure from spectropolarimetric analysis. We therefore conclude that the occulting material producing the dips in the light curve must be located close to the corotation radius in the circumstellar disk. 

    We thus defined the rotational ephemeris so that the rotational phase 0.5 corresponds to the photometric minimum, and maximum occultation of the central star by circumstellar dust. Furthermore, the stability of the K2 light curve, with recurrent dips occurring over more than 18 rotational cycles, is reminiscent of the behavior of the prototype of dippers, AA Tau. This suggests that the dips are caused by an inner disk warp interacting with the stellar magnetopshere close to the corotation radius.    
      
    V807 Tau's spectrum exhibits  \ha, \hb, and \hg \ emission lines typical of accreting TTs, from which we deduce a mass accretion rate of $\sim$~10$^{-9}$ \msunyr. Indeed the double-peaked shape of the Balmer lines with a central absorption component is quite similar to the line profiles seen in other dippers, such as AA Tau \citep{Bouvier03, Bouvier07a} and LkCa15 \citep{Alencar18}, which also exhibit moderate mass accretion rates. 
    We find that the intensity of the redshifted wing of the \hb \ and \hg \ line profiles, up to a velocity of about +300 \kms, is modulated at the stellar rotation period. This signature is consistent with  ballistic accretion along funnel flows connecting the inner disk to the stellar surface, which periodically cross the line of sight.  
    
    In addition, the Balmer line profiles provide evidence consistent with a magnetospheric inflation scenario. 
    The last 3 observations exhibit depressed blueshifted wings,  which may indicate an episodic outflow. 
    Magnetospheric inflation occurs when the magnetic field lines connect to the disk at a radius different from the corotation radius. 
    In such a case, the lower and upper tips of the accretion funnel flow rotate at a different period. 
    This induces a torque that inflates the field lines until they open up and reconnect, producing an ejection \citep{Goodson97}.
    The simulations of \cite{Zanni13} revealed that this phenomena is periodic, on a timescale amounting to several rotation period. 
    This accounts for the fact that these episodic ejection signatures are rarely seen in monitoring campaigns that cover only a few rotational periods \citep[see, e.g.,][]{Pouilly20}.
    Furthermore, the Gaussian decomposition of the H$\alpha$ line profile reveals a linear correlation between the velocities of the blueshifted and redshifted absorption components (see Sect. \ref{subsec:balmer}), which has previously been interpreted as a signature of a magnetospheric inflation cycle \citep{Bouvier03}.
    
    Accretion funnel flows are expected to dissipate the kinetic energy of the accreted material at the stellar surface, producing an accretion shock, observable as a localized bright spot on the stellar photosphere.
    The narrow component of the He I 587.6 nm line is thought to arise in the post-shock region \citep{Beristain01}. We find both the intensity and the radial velocity of the HeI narrow component to be rotationally modulated at the stellar rotation period, consistent with this interpretation. 
    A detailed modeling of the He~I NC radial velocity curve indicates that the hot spot faces the observer at phase 0.55, at a latitude of $\sim$70$^\circ$.
    This is consistent with the photometric minimum, which indicates that the structure that occults the star is located at the same azimuth as the accretion shock.
    From this analysis we also derive a downward radial velocity of +9.5 \kms\ for the gas in the postshock region. 
    
    The Doppler and Zeeman-Doppler analyzes support the location we derived above for the accretion shock at the stellar surface.  
    The Doppler map yields a bright area extending between phase 0.5 and 0.8, at a latitude ranging from 50$^\circ$ and 75$^\circ$.
    The ZDI analysis reveals a mostly poloidal magnetic topology, with a significant dipolar component whose pole is anchored at phase 0.7 and a latitude of 50$^\circ$ on the stellar surface. The hot spot that signals the accretion shock at the stellar surface thus seems to be spatially associated with the magnetic pole of the large-scale magnetospheric component interacting with the disk at a distance of a few stellar radii. We note, however, that redshifted absorptions reaching below the continuum are more clearly seen in the \hg\ line profile at phases 0.26 and 0.34, which suggests multiple accretion funnel flows onto the star. 
    
    From the Stokes V profile of He~I NC component, which probes the strength of the magnetic field at the foot of the funnel flow, we derive a value of the longitudinal magnetic field of 2 kG at the pole.
    However, we cannot derive the magnetic topology from this line profile, which would allow us to estimate the strength of the large-scale dipolar component that interacts with the inner disk. 
    Instead, since 
     the periodicity of the light curve dips is consistent with the rotation period of the star, we may assume that 
     the inner dusty disk is truncated close to corotation. 
    We thus assume that the magnetospheric truncation radius  \rmag\ $\simeq$ \rcor\ = 5.8 $\pm$ 0.4 \rstar, and derive the required strength of the  corresponding dipolar magnetic field from the \cite{Bessolaz08} expression:
    \begin{equation} 
    {\frac{r_{mag}}{R_{\star}}}  = 2 m_s^{2/7} B_{\star}^{4/7} \dot{M}_{acc}^{-2/7} M_{\star}^{-1/7} R_{\star}^{5/7}, 
    \end{equation}
    where m$_s\approx1$, B$_{\star}$ is the equatorial magnetic field strength (\textit{i.e.,} half of the polar value) in units of 140 G, \macc\ is the mass accretion rate in units of 10$^{-8}$~\msunyr, \mstar\ the stellar mass in units of 0.8~\msun, and \rstar\ the stellar radius in units of 2~\rsun.
    This yields B$_{dipole}$ $\sim$ 550 G, a value consistent  with the B$_l$ we measured in the post shock region, which includes the octupolar component as well.

    The photometric and spectroscopic properties and variability of V807 Tau, as well as its magnetic topology, are all reminiscent of other dippers, where the periodic occultation of the central star results from a rotating inner disk warp. Most such dippers, however, are found to be seen at moderate to high inclinations \citep{Mcginnis15}, a necessary condition for the inner disk warp to intersect the line of sight to the star \citep{Bodman17}. This does not seem to be the case for V807 Tau.  
    From the rotation period, \vsini, and the stellar radius, we derived an inclination of 41 $\pm$ 10$^\circ$, while the DI analysis yielded an independent estimate of i = 53 $\pm$ 5$^\circ$. The moderate inclination of the system probably accounts for the fact that we do not see as deep redshifted absorptions, reaching below the continuum, as seen in the line profiles of other, more inclined dippers \citep[see][for an example]{Alencar18}. It may also explain why the dips in V807 Tau's light curve are relatively shallow, amounting to only about 0.1 mag, compared to other, more inclined systems (e.g., 1.5 mag dips in AA Tau  seen at an inclination of $\sim$75\degr, \cite{Bouvier99}). Whether the inner disk warp induced by the inclined magnetosphere can reach high enough above the disk midplane so as to produce shallow stellar occultations in a system seen at moderate inclination has yet to be established. 
    
    Following the equation of \cite{Monnier02} the dust sublimation radius, $r_{sub}$, is given by: $$r_{sub}=(\sqrt{Q_r}/2)\times(T_{eff}/T_s)^{2}$$, where $Q_r = 1$ is the ratio of the dust absorption efficiencies, and $T_s = 1500$ K is the dust sublimation temperature. We thus derive a dust sublimation radius $r_{sub}=$ 3.7 $\pm$ 0.5 \rstar,    
    consistent with the presence of dust well within the corotation radius. In such a circumstance, dust may survive within the funnel flow, thus perhaps reaching high enough above the disk plane so as to intercept the line of sight. This explanation was put forward to account for shallow dippers in NGC 2264 \citep{Stauffer15}. The possibility of dust survival in accretion funnel flows was further investigated from dust sublimation models developed by   \cite{Nagel20}, which might apply to V807 Tau.

\section{Conclusions}
\label{sec:conclusion}
    V807 Tau was singled out as a stable periodic dipper from its Kepler K2 light curve. The aim of the CFHT/ESPaDOnS follow-up campaign was to investigate the origin of the variability of the system by coupling  spectropolarimetry and photometry. Specifically, 
    the aim of this study was to characterize the magnetic properties and accretion process in this system. 

    We report several clues to the magnetospheric accretion process being at work in V807 Tau. 
    The redshifted wings of the H$\beta$ and H$\gamma$ line profiles are modulated on the stellar rotation period, which is a signature of accretion funnel flow passing on the line of sight.
    The spectrum of V807 Tau exhibits a narrow component of the He I line at 587.6 nm, indicative of an accretion shock, whose visibility is also modulated at the stellar rotation period. 
    Furthermore, its intensity 
    is maximum at the phase where we measure the largest value of the longitudinal component of the magnetic field.  
    Finally, the Zeeman-Doppler analysis of Stokes I and V profiles reveals a bright structure that supports the location of the accretion shock at the stellar surface and yields 
    and a magnetic field topology showing an important a dipolar component. All these signatures are typical of low-mass CTTSs and are thought to arise from magnetically controlled accretion between the inner disk and the star. Strong variability is also seen in the blue wing of the Balmer line profiles, which suggests variable outflows. These, however, are not modulated at the stellar rotation period and appear to be transient, perhaps the result of magnetospheric ejections. 

    
    While an inner disk warp is expected to result from the interaction of the inner disk with the inclined magnetosphere at a distance of a few stellar radii, independent estimates suggest that the system is seen at a moderate inclination of about 50 $\degr$, which is unusually low for a dipper. The low luminosity of the system, its moderate rotation, and low mass accretion rate, translate into the sublimation radius lying well inside the corotation and truncation radii. We suggest that in this specific configuration, dust may survive along a significant portion of the accretion funnel flow, thus giving rise to shallow dips in moderately inclined systems.    
    

\begin{acknowledgements}
      We thank an anonymous referee whose comments improved the content of this paper. This project has received funding from the European Research Council (ERC) under the European Union’s Horizon 2020 research and innovation programme (grant agreement No 742095 ; SPIDI : Star-Planets-Inner Disk- Interactions; http://www.spidi-eu.org).
      This paper includes data collected by the Kepler mission. Funding for the Kepler mission is provided by the NASA Science Mission directorate.
      This work has made use of the VALD database, operated at Uppsala University, the Institute of Astronomy RAS in Moscow, and the University of Vienna. 
      This project was funded in part by INSU/CNRS Programme National de Physique Stellaire and Observatoire de Grenoble Labex OSUG2020.
      S.H.P. Alencar acknowledges financial support from CNPq, CAPES and Fapemig.
      K. Grankin acknowledges the partial support from the Ministry of Science and Higher Education of the Russian Federation (grant 075-15-2020-780).
\end{acknowledgements}


\begin{appendix}

\section{K2-light curve's decomposition and dips structure analysis}
\label{ap:v807lc}

\begin{figure*}
    \centering
    \includegraphics[width=.7\textwidth]{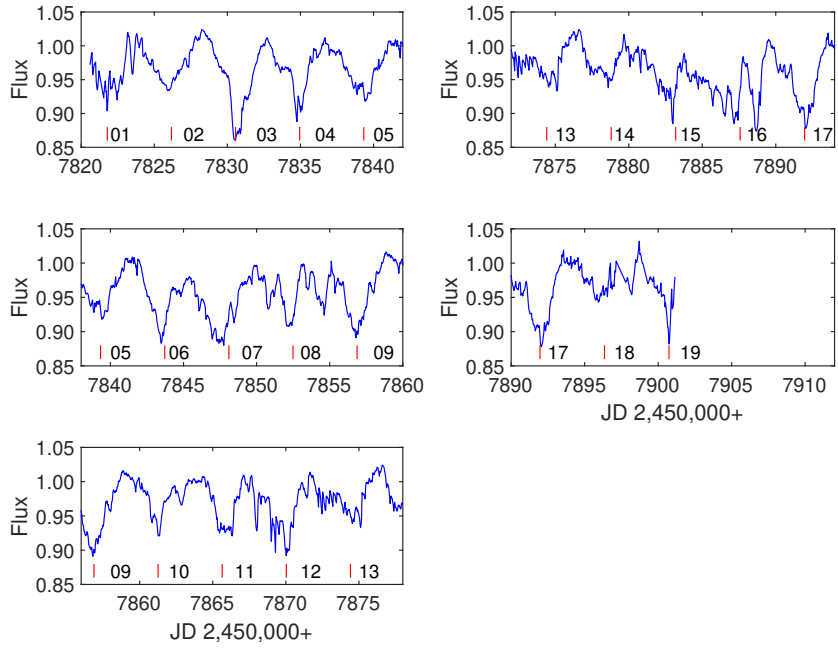}
    \caption{K2-light curve of V807 Tau split into sections of five successive minima. Each panel overlaps the previous one by one minimum. The red lines identify the location of the minima, separated by a constant time interval corresponding to the rotational period of the star. }
    \label{fig:v807lc}
\end{figure*}

Fig.~\ref{fig:v807lc} presents the K2-light curve split in sections consisting of five successive dips, except for the last panel. Each section overlaps with the next one by one minimum, and each minimum is numbered.  

\begin{figure*}
    \centering
    \includegraphics[width=.7\textwidth]{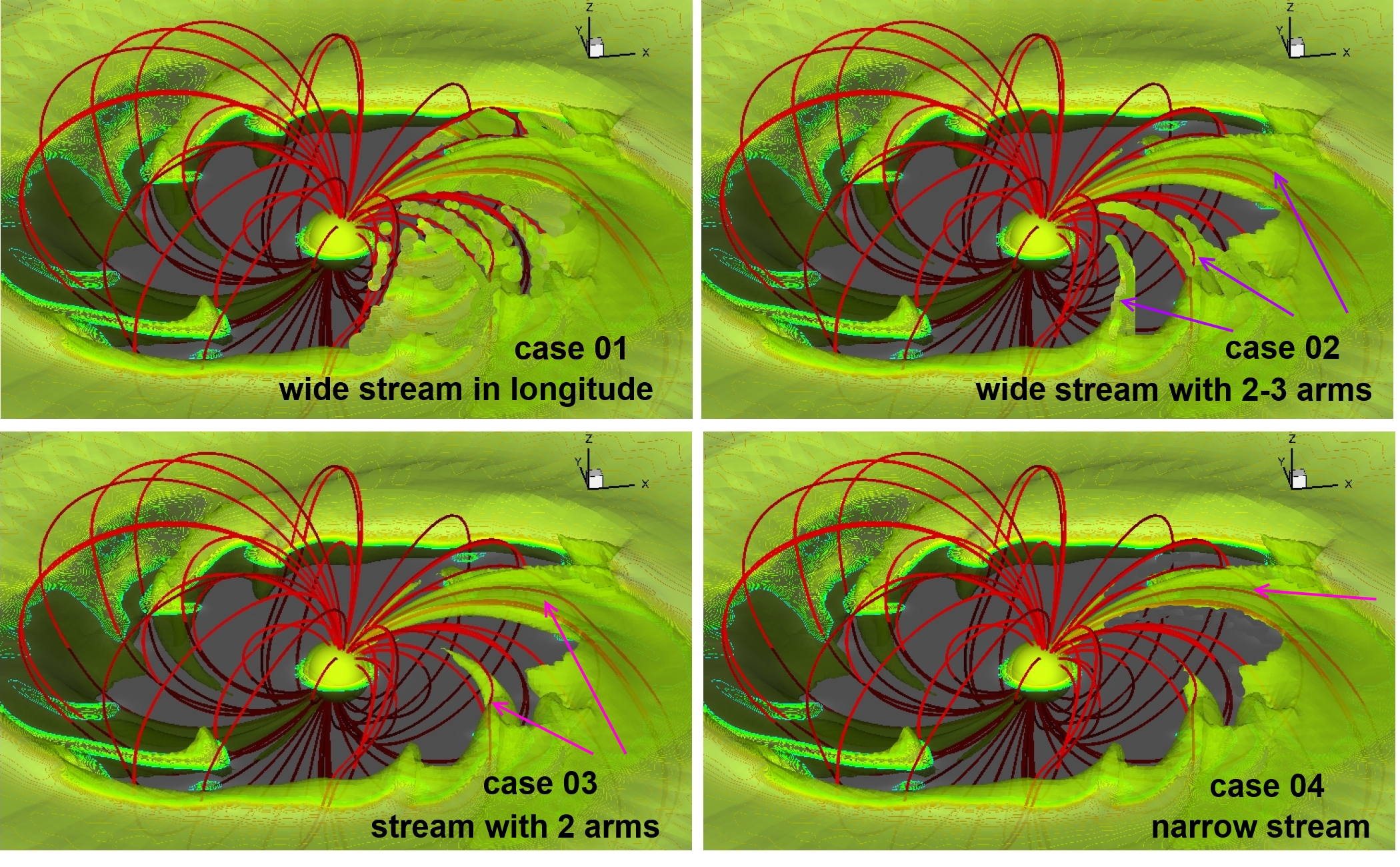}
    \caption{Artist view adapted from the 3D simulations of \cite{Romanova13} of accretion funnel flows along an inclined magnetosphere. Those describe the four cases invoked in Appendix \ref{ap:v807lc} to explain the different dips shapes. The top left panel represents an accretion flow covering a wide range in longitude (case 1), the top right panel shows a wide accretion flow split in several arms (case 2), the bottom left panel represents an accretion flow narrower than cases 1 and 2 and split in two different arms (case 3), and the bottom right panel shows a typical accretion flow of one narrow stream (case 4).}
    \label{fig:visuAccCol}
\end{figure*}

We notice the variety of depth, shape, and duration of the minima and we propose here different hypotheses to account for them, focusing on the inner disk structure and magnetospheric topology. 
We highlight four different cases illustrated by the simulations shown in Fig.~\ref{fig:visuAccCol} adapted from  \cite{Romanova13} : i) Shallow and wide minima (\#2, 5, 11, and 14) that can be attributed to a wide stream spread in azimuth (case 01).
ii) Series of narrow dips with variable depth (\#12, 16) that are consistent with a wide accretion flow split in several arms near the truncation radius (case 02).
iii) Shallow and narrow dips that are preceded by another
decrease in brightness with a 0.3-0.4 phase shift relative to the main minimum (\#8, 9). 
These may be explained by a 2-arm stream near the truncation radius (case 03). 
iv) Single narrow dips (\#3, 4, 6, 17, and 19), which would correspond to a narrow accretion funnel flow (case 04).

We also notice temporal decreases in the maximum brightness level that can last for several cycles (for example, from dips \#6 to 9, and from \#14 to 17), which suggests that   the star may remain partly obscured between major dips. 





\section{Gaussian decomposition of the H$\alpha$ line profile.}
\label{ap:GaussHa}

\begin{figure*}
    \centering
    \includegraphics[width=.95\textwidth]{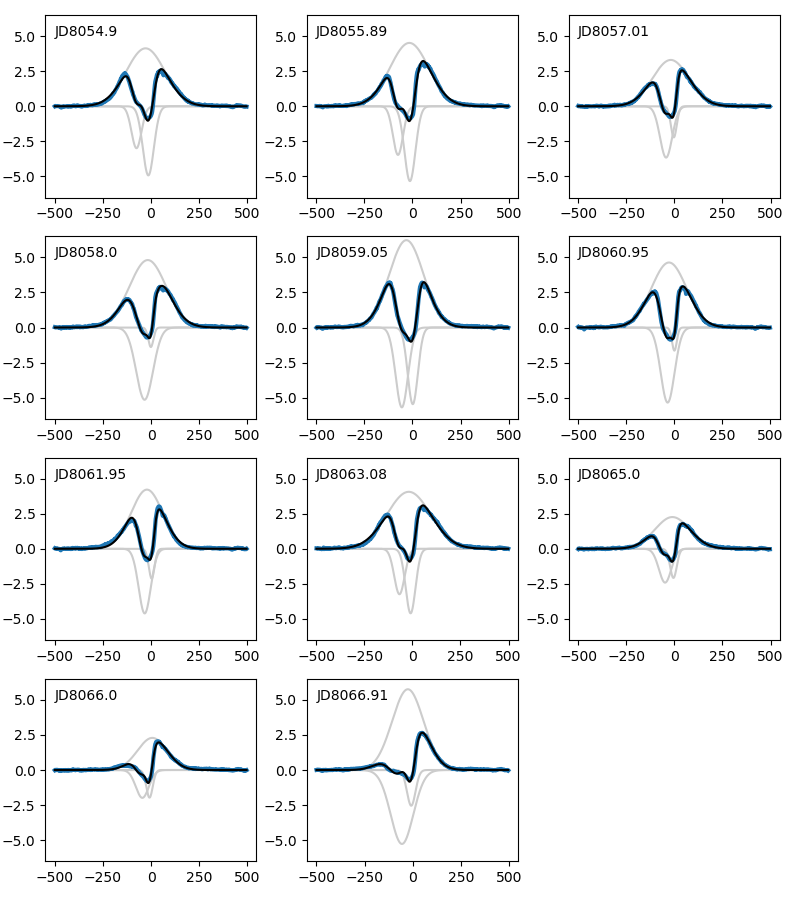}
    \caption{Gaussian decomposition of the H$\alpha$ line profile. The observed profiles are shown in blue, the fit is shown in black, and the different components are detailed in gray.}
    \label{fig:HaDecomp}
\end{figure*}

Figure \ref{fig:HaDecomp} illustrates examples of the Gaussian decomposition of the H$\alpha$ line profile, consisting of a broad central emission and two narrow absorption components. 

\end{appendix}


\bibliographystyle{aa}
\bibliography{bib}

\end{document}